\documentclass[%
 reprint,
 amsmath,amssymb,
 aps,
prb,
]{revtex4-2}

\usepackage{graphicx}
\usepackage{dcolumn}
\usepackage{bm}
\usepackage{xcolor}
\usepackage{soul}

\begin{document}

\preprint{APS/123-QED}

\title{Nonequilibrium Green's Function Formalism\\
Applicable to Discrete Impurities in Semiconductor Nanostructures}

\author{Nobuyuki Sano}
 \email{sano.nobuyuki.gw@u.tsukuba.ac.jp}
\affiliation{%
 Institute of Applied Physics, University of Tsukuba \\
 1-1-1 Tennoudai, Tsukuba, Ibaraki 305-8573, Japan 
}%




\date{\today}

\begin{abstract}
A new theoretical framework for the nonequilibrium Green's function (NEGF) scheme is presented to account for the discrete nature of impurities doped in semiconductor nanostructures. The short-range part of impurity potential is included as scattering potential in the self-energy due to {\em spatially localized} impurity scattering, and the long-range part of impurity potential is treated as the self-consistent Hartree potential by coupling with the Poisson equation. The {\em position-dependent} impurity scattering rate {\em under inhomogeneous impurity profiles} is systematically derived so that its physical meaning is clarified. The position dependence of the scattering rate turns out that it is represented by the `center of mass' coordinates in the Wigner coordinates, rather than the real-space coordinates. 
Consequently, impurity scattering is intrinsically nonlocal in space. 
The proposed framework is applied to cylindrical thin wires under the quasi-one-dimensional (quasi-1D) approximation. We show explicitly how the discrete nature of impurities affects the transport properties such as electrostatic potential, local density of states, carrier density, scattering rates, and mobility. 
\end{abstract}

\maketitle


\section{\label{sec:intro}Introduction}
Quantum electron transport with impurity scattering in doped semiconductors has been often studied by the NEGF scheme, in which the screened Coulomb potential induced by ionized impurities (and carriers) is incorporated into the self-energy due to impurity scattering~\cite{Lake1992,Lake1997,Haug2008,Oh2008,Oh2013}. However, the self-energy is always assumed to be the average of all possible spatial configurations of impurities within a semiconductor substrate~\cite{Kohn1957,Abrikosov1963,Doniach1974,Mahan2000}. This averaging is sometimes referred to as `self-averaging' because the device substrate is assumed to be much larger than the phase coherence length, so the substrate could be regarded as a collection of many uncorrelated pieces. 
This leads to the realization of ensemble-averaged transport properties even under the given doped material. Namely, the transport properties do not depend on the microscopic details of impurity distributions inside the substrate. 

The approaches mentioned above are, however, valid only in cases where the spatial variations of physical quantities such as electrostatic potential and electron density are gradual, compared with the phase coherence length. 
Hence, these approaches break down in nanostructures where the substrate is too small to impose self-averaging. In fact, electron transport properties in a nanostructured semiconductor often show large variabilities even at room temperature~\cite{Bjork2009}, in which impurity configurations inside the substrate are expected to be crucial. 
This is actually one of the most serious problems encountered in leading-edge Si-based semiconductor devices; the device characteristics such as threshold voltage greatly vary from device to device~\cite{Hoeneisen1972,Keyes1975,Wong1993,Sano2000,Miranda2012}.  

The physical origin of such variabilities in transport properties under nanostructures lies in the discrete nature of impurities doped in a semiconductor substrate: The `long-range' part of each impurity potential leads to potential fluctuations whose wavelength is longer than the screening length~\cite{Sano2002,Sano2017sispad,Sano2020}, and the `short-range' part of impurity potential leads to the `position-dependent' impurity scattering~\cite{Sano2021,Sano2021pre}.  
It should be noted that such `short-range' scattering cannot be ignored even in ultra-scaled electron devices because of the fundamental requirement from the transport equation leading to the law of increasing entropy~\cite{Sano2004PRL}. 
Therefore, it is of crucial importance to take into account the discreteness of impurities properly in both the `long-range' potential fluctuations and the `short-range' impurity scattering for any reliable theoretical analyses of electron transport in doped semiconductors.

In the past few decades, this problem has been studied by another method based on the NEGF~\cite{Martinez2007,Bescond2009,Akhavan2012,Carrillo2014}, in which the usage of self-energy due to impurity scattering has been intentionally avoided. Instead, discrete impurities represented by point charges are treated in an unperturbed way; they are directly introduced into both the unperturbed Hamiltonian in the retarded Green's function and the Poisson equation. This approach is intuitively appealing. However, the impurity potential induced by a point charge is intrinsically singular at the impurity's site and, thus, appropriate treatments of such singular potential would be complicated especially in the finite difference method or similar ones. 
Indeed, such NEGF simulations often lead to numerical instabilities~\cite{Lee2024}. 
These problems might be, however, resolved by treating the doped impurities with the first-principles approaches in which the density functional theory is combined with the NEGF~\cite{Markussen2007,Rurali2008,Wilson2019}. Yet, because of its computational burden, employing this approach to study issues such as variabilities in transport characteristics would still be difficult. 

In the present study, we take a traditional approach under the NEGF framework: The discrete nature of doped impurities in a semiconductor substrate is incorporated into the self-energy due to impurity scattering. 
To achieve this goal, we need to revisit a few critical approximations included explicitly or implicitly in the traditional approaches. 

First, self-averaging imposed on impurity scattering is removed. Since the locations of the scattering events in space are washed out after self-averaging, we need to construct a new expression of the self-energy due to impurity scattering so that the locations of impurities are directly reflected in the scattering processes. 
Second, potential fluctuations induced by ionized impurities in the substrate are introduced as the self-consistent Hartree potential by coupling the NEGF scheme with the Poisson equation. We would like to stress that such `long-range' potential fluctuations are completely ignored in the traditional approaches because of the assumption of complete screening and/or self-averaging, which are unrealistic under nanostructures~\cite{Sano2018}. 
Third, under inhomogeneous impurity-density profiles, the (global) position dependence in the impurity scattering rate should be systematically derived. In the traditional approaches, the impurity scattering rate derived from Fermi's golden rule, which is valid only under homogeneous structures, is employed and its (global) position dependence is in most cases introduced on an {\it ad hoc} empirical basis~\cite{Jacoboni1989,Vasileska2006}. 
The derivation of the self-energy due to impurity scattering reveals that the position dependence is actually represented by the Wigner coordinates, rather than the ordinary real-space coordinates. As a consequence, the self-energy matrix of impurity scattering becomes intrinsically non-diagonal in the real-space representation. 

The present paper is organized as follows. 
In Sec.~\ref{sec:framework}, a generic theoretical framework of the NEGF scheme coupled with the Poisson equation is presented. The physical interpretation of splitting the `long-range' and `short-range' parts of impurity potential is also explained. To apply the present framework to a cylidrical wire structure, the quasi-1D approximation employed for the NEGF calculations is explained in Sec.~\ref{sec:quasi-1D}.  In Sec.~\ref{sec:R&D}, the calculation results and discussion are presented. Finally, conclusions are drawn in Sec.~\ref{sec:concl}.

\section{Generic framework of NEGF under discrete impurities}
\label{sec:framework}
The present theoretical framework consists of the Poisson equation for finding the electrostatic potential and the NEGF for the electron density. Both are coupled to each other to account for the discrete nature of impurities in a self-consistent manner. Here, we describe a new theoretical scheme to account for this point. 

\subsection{Electrostatic potential under discrete impurities}
\label{sec:potential}
We suppose that the semiconductor substrate is doped by donor impurities. The electrostatic potential of impurities in a substrate is split into the long- and short-range components so that each component satisfies the corresponding Poisson equations. In the following, we outline how the potential could be split and its physical implications.

The Poisson equation for the `long-range' components of the potential under the given configuration of impurities ${\left\{ {{{\mathbf{R}}_{i}}} \right\}}$, where ${\left\{ {{{\mathbf{R}}_{i}}} \right\}}$ represents a set of the position vectors of each impurity, is assumed to be given by
\begin{align}
\nabla  \cdot \left( {{\varepsilon _s}\nabla {V_{long}}} \right) =  - e \left[ {N_{d,long}^ + \left( {{\mathbf{r}};\left\{ {{{\mathbf{R}}_{i}}} \right\}} \right) - n\left( {\mathbf{r}} \right)} \right] ,
\label{eq:Poissonlong}
\end{align}
where $\varepsilon _s$ is the `static' dielectric constant of the semiconductor substrate, $e~(>0)$ is the elementary charge and $n$ is the electron density obtained from the transport equation, namely, the Kadanoff-Baym-Keldysh equation for the lesser Green's function~\cite{Kadanoff1962,Keldysh1965}. $N_{d,long}^ +$ is the effective donor density derived by the `long-range' components of the potential due to discrete impurities, which are coupled with the Poisson equation for the `short-range' components of the potential in a self-consistent manner, as described below. 
Notice that Eq.~\eqref{eq:Poissonlong} reduces to the Poisson equation in the conventional device simulations by replacing $N_{d,long}^ + \left( {{\mathbf{r}};\left\{ {{{\mathbf{R}}_{i}}} \right\}} \right)$ with the `jellium' donor density $\bar N_{d}^ + \left( {{\mathbf{r}}} \right)$. 
Hence, $V_{long}$ in Eq.~\eqref{eq:Poissonlong} represents the `long-range' Hartree potential in the present framework. 

We now determine the unknown effective donor density $N_{d,long}^ + $.
Under the homogeneous bulk structures in thermal equilibrium, $V_{long}$ becomes nearly a flat potential in space because each impurity is fully screened and, thus, the right-hand side of Eq.~\eqref{eq:Poissonlong} must vanish. In other words, $N_{d,long}^ + $ must balance with the electron density in equilibrium due to the (global) charge neutrality condition. Hence, electron density becomes inhomogeneous in space even under bulk structures because of the (local) accumulation of electrons around ionized donor impurities for screening. As a result, $N_{d,long}^ +$ may be approximated by 
\begin{align}
N_{d,long}^ + \left( {{\mathbf{r}};\left\{ {{{\mathbf{R}}_i}} \right\}} \right) \approx \sum\limits_{l = 1}^{N_d^{tot}} {{n^{eq}}\left( {{\mathbf{r}} - {{\mathbf{R}}_l}} \right)} , 
\label{eq:Ndlong}
\end{align}
where $n^{eq}\left( {{\mathbf{r}} - {{\mathbf{R}}_l}} \right)$ is the electron density {\em in equilibrium} induced by a donor impurity at $\mathbf{R}_l$, and ${N_d^{tot}}$ is the total number of donor impurities in the substrate. 
From the above considerations, $n^{eq}$ must fulfill the Poisson equation for the (short-range) screening potential $v_{sc}$ due to a single impurity;
\begin{align}
\nabla  \cdot \left( {{\varepsilon _{s}}\nabla {v_{sc}}\left( {\mathbf{r}} \right)} \right) =  - e\left[ {\delta \left( {\mathbf{r}} \right) - n^{eq} \left( {\mathbf{r}} \right)} \right] ,
\label{eq:nd+}
\end{align}
where an impurity is assumed to be a point charge.  Here, we have also assumed that the mean separation of impurities is larger than the screening length so that screening electrons of different impurities do not overlap with each other. 
{
Throughout the present numerical calculations, we assume the material Si, in which the mean separation is always larger than the screening length in the ranges of donor density considered in the present article. However, the overlap of multiple impurities could be significant in other materials such as some compound semiconductors. In this case, this assumption cannot be justified in a strict sense. Within the present framework, however, the effective donor density could be treated as an adjustable functional and, thus, by including the higher-order corrections beyond the Born approximation in the self-energy due to impurity scattering, 
Eq.~\eqref{eq:nd+} might still be used.} 

Equation~\eqref{eq:nd+} needs to be solved {\em under equilibrium} with appropriate boundary conditions consistent with the device structure. 
The short-range part of the electrostatic potential in the substrate is then described by 
\begin{align}
V_{short} \left( {{\mathbf{r}};\left\{ {{{\mathbf{R}}_i}} \right\}} \right) = \sum\limits_{l = 1}^{N_d^{tot}} {{v_{sc}}\left( {{\mathbf{r}} - {{\mathbf{R}}_l}} \right)} ,
\label{eq:Vshort}
\end{align}
and it satisfies the Poisson equation for the `short-range' part of the potential given by 
\begin{align}
\nabla  \cdot \left( {{\varepsilon _s}\nabla {V_{short}}} \right) =  - e 
\left[
\sum\limits_{l = 1}^{N_d^{tot}} 
 {\delta \left( {\mathbf{r}} - {\mathbf{R}_l} \right)} 
- N_{d,long}^ + \left( {{\mathbf{r}};\left\{ {{{\mathbf{R}}_{i}}} \right\}} \right) 
\right].
\label{eq:Poissonshort}
\end{align}
Equations~\eqref{eq:Poissonlong} and \eqref{eq:Poissonshort} describe the electrostatic potential in a semiconductor substrate under the given distribution of impurities. 
Combining these two equations results in the ordinary Poisson equation commonly used in device simulations, in which discrete impurities are represented by point charges; 
\begin{align}
\nabla  \cdot \left( {{\varepsilon _s}\nabla {\left( V_{long}+V_{short} \right)}} \right) =  - e 
\left[
\sum\limits_{l = 1}^{N_d^{tot}} 
 {\delta \left( {\mathbf{r}} - {\mathbf{R}_l} \right)} 
- n\left( {\mathbf{r}} \right)
\right].
\label{eq:Poissonfull}
\end{align}

In the case of bulk structures where the effects associated with the boundary and/or interface of the substrate are safely ignored, Eq.~\eqref{eq:nd+} is solved along with $n^{eq}$ obtained under the linear approximation, and $v_{sc}$ is given by the usual Yukawa potential,
\begin{align}
{v_{sc}}\left( {\mathbf{r}} \right) = \frac{e}{{4\pi {\varepsilon _s}}}\frac{{{e^{ - {q_c}\left| {\mathbf{r}} \right|}}}}{{\left| {\mathbf{r}} \right|}} .
\label{eq:Yukawa}
\end{align}
The effective donor density is then given by
\begin{align}
N_{d,long}^ + \left( {{\mathbf{r}};\left\{ {{{\mathbf{R}}_{i}}} \right\}} \right) = \sum\limits_{l = 1}^{N_d^{tot}} {\frac{{{q_c}^2}}{{4\pi }}\frac{{{e^{ - {q_c}\left| {{\mathbf{r}} - {{\mathbf{R}}_{l}}} \right|}}}}{{\left| {{\mathbf{r}} - {{\mathbf{R}}_{l}}} \right|}}} ,
\label{eq:Nd_long}
\end{align}
where $q_c$ is the inverse of the screening length at {\em each position of impurity}. 
Notice that $q_c$ is determined by Eq.~\eqref{eq:nd+} and, thus, fixed by the {\em equilibrium} electron density $n^{eq}\left( {\mathbf{r}} \right)$. 
Also, Eq.~\eqref{eq:Nd_long} implies that the charge density of each impurity spreads over the substrate, rather than a point charge. This simply reflects the fact that the potential in Eq.~\eqref{eq:Poissonlong} is the long-ranged whose length-scale is greater than the screening length.  
This is consistent with the scheme we have proposed to implement the discrete impurities in the conventional device simulations~\cite{Sano2000,Sano2002}.
For more complicated device structures where the boundary and/or interface of the device substrate crucially affect the electrostatic potential, Eq.~\eqref{eq:nd+} must be solved with proper boundary conditions. This can be done by using the Green's function with image charges, as we have demonstrated in~\cite{Sano2018}. 

We would like to comment on the physical interpretation of Eqs.~\eqref{eq:Poissonlong} and \eqref{eq:Poissonshort}. 
Within the present framework, the short-range part of impurity potential in Eq.~\eqref{eq:Poissonshort} is always represented by fully screened scattering potential, irrespective of actual electron density around impurities. Hence, the `long-range' components, greater than screening length, are always treated as the self-consistent Hartree potential and, thus, included in the unperturbed Hamiltonian. 
When impurities are fully- or over-screened, the `long-range' components of impurity potential are completely masked. Then, the Hartree potential obtained from Eq.~\eqref{eq:Poissonlong} becomes nearly flat, and an implicit assumption of taking the long-wavelength limit in the traditional method with the `jellium' impurity 
is justified.  
On the other hand, when screening is incomplete due to, say, an insufficient number of electrons in the depletion regions, unscreened components of impurity potential show up by Eq.~\eqref{eq:Poissonlong} as potential fluctuations, and they are accounted for as the Hartree potential. 
Since the {\em collective response} (screening) of electrons against excess or deficient charge distributions is induced by the `long-range' components of impurity potential, the present approach seems to be physically more consistent than the traditional ones where those components are either fully ignored 
or taken into account as stochastic scattering potential. 

\subsection{NEGF under discrete impurities}
\label{sec:general}
%
%
As for the NEGF framework, we employ the usual partition method~\cite{Caroli1971,Datta2005}; the device region is connected to the semi-infinite leads at the left and right ends of the device. The left and right leads are 
{
non-doped and}
in thermal contact with the (heat and particle) reservoir described by the equilibrium Fermi-Dirac distributions at room temperature ($T=300$ K). 
Then, the NEGF scheme consists of two independent Green's functions of the {\em device region}; the retarded Green's function and the lesser Green's (correlation) function. 
We introduce the discrete nature of impurities into both Green's functions in a way to be consistent with the long- and short-range Poisson equations described in Sec.~\ref{sec:potential}.

The retarded Green's function in the real-space representation under steady state is given by 
\begin{align}
{G^r} \left( {{\mathbf{r}},{\mathbf{r'}};E} \right) & = \left\langle {\mathbf{r}} \right|{{\hat G}^r}\left( E \right)\left| {{\mathbf{r'}}} \right\rangle   \nonumber \\
& = \left\langle {\mathbf{r}} \right|\frac{1}{{E + i{0^ + } - {{\hat H}_D} - {{\hat \Sigma }^r}_{L/R} - {{\hat \Sigma }^r}_{imp}}}\left| {{\mathbf{r'}}} \right\rangle .
\label{eq:Gr}
\end{align}
where ${{{\hat H}_D}}$ is the unperturbed {\em single-particle} Hamiltonian of the device region and it is, under the effective mass approximation, expressed by 
\begin{align}
{{\hat H}_D} = \frac{{{{{\mathbf{\hat p}}}^2}}}{{2m}} -e {V_{long}}\left( {{\mathbf{\hat r}};\left\{ {{{\mathbf{R}}_{i}}} \right\}} \right) .
\label{eq:H0}
\end{align}
Here, ${\mathbf{\hat p}}$ is the momentum operator of electron, $m$ is the effective mass, and $V_{long}$ is the electrostatic potential (operator) obtained from Eq.~\eqref{eq:Poissonlong} under the given distribution of impurities, $\left\{ {{{\mathbf{R}}_{i}}} \right\}$. 
The self-energy operator for the contact with the left and right leads is denoted by $\hat \Sigma _{L/R}^\alpha$,  
and ${{{\hat \Sigma }^\alpha}_{imp}}$ denotes the self-energy due to impurity scattering inside the device region. Here, $\alpha = r$ or $<$. 

The lesser Green's function under steady state is given by 
\begin{align}
{G^ < }\left( {{\mathbf{r}},{\mathbf{r'}};E} \right) = \left\langle {\mathbf{r}} \right|{{\hat G}^r}\left( {{{\hat \Sigma }^<}_{L/R} + {{\hat \Sigma }^<}_{imp}} \right){{\hat G}^a}\left| {{\mathbf{r'}}} \right\rangle  ,
\label{eq:G<}
\end{align}
where ${{\hat G}^a}$ is the advanced Green operator and given by 
${{\hat G}^a} =  {\hat G}^{r\dag} $. 
Then, the electron density $n \left( {\mathbf{r}} \right) $ in Eq.~\eqref{eq:Poissonlong} is calculated by 
\begin{align}
n\left( {\mathbf{r}} \right) =  - i\int {\frac{{dE}}{{2\pi }}{G^ < }\left( {{\mathbf{r}},{\mathbf{r}};E} \right)} .
\label{eq:nr}
\end{align}
%

%
The self-energy operator due to the (left and right) contacts is given by 
\begin{align}
{{\hat \Sigma }^\alpha}_{L/R} \left( E \right)= {{\hat \Sigma }^\alpha}_{L}\left( E \right) +{{\hat \Sigma }^\alpha}_{R}\left( E \right), 
\label{eq:sigma_LR}
\end{align}
where $\hat \Sigma _{L(R)}^\alpha$ is self-energy operator due to the contact with the left (right) lead and given by the usual expression derived by the partition method:
\begin{align}
\hat \Sigma _{L(R)}^\alpha  \left( E \right) = \hat V_{L(R)}^\dag {{\hat g}_{L(R)}}^\alpha \left( E \right)\hat V_{L(R)}. 
\label{eq:sigma_LR2}
\end{align}
Here, $\hat V_{L(R)}$ is the interaction potential (operator) between the left (right) lead and the device region, and ${{\hat g}_{L(R)}}^\alpha$ is the unperturbed Green's operator of the semi-infinite left (right) lead. 
In the real-space representation, $\hat V_{L(R)}$ has nonzero elements only at the sites adjacent to the device region. Hence, only a small number of elements of ${{\hat g}_{L(R)}}^r$ is required to calculate Eq.~\eqref{eq:sigma_LR2} and they are found by the recursive method~\cite{Caroli1971,Datta2005}. 
Assuming that the left and right leads are in thermal equilibrium with the reservoirs, 
${{\hat g}_{L(R)}}^<$ is given by 
\begin{align}
\hat g_{L(R)}^ < \left( E \right) = if_{FD}^{L(R)}\left( E \right)\left\{ { - 2\operatorname{Im} \left[ {\hat g_{L(R)}^r\left( E \right)} \right]} \right\} ,
\label{eq:g<LR}
\end{align}
where $f_{FD}^{L(R)} \left( E \right)$ is the Fermi-Dirac distribution of the left (right) reservoir with the chemical potential $\mu _{L(R)}$. 

\subsection{Self-energy due to impurity scattering}
\label{sec:sigma_imp}
The self-energy ${{{\hat \Sigma }^\alpha}_{imp}}$  due to impurity scattering under discrete impurities, which is one of the main concerns in the present study, is treated by the self-consistent Born approximation. 
Then, ${{{\hat \Sigma }^\alpha}_{imp}}$ is given by 
\begin{align}
{{\hat \Sigma }^\alpha }_{imp} = e^2 {V_{short}}{\left( {{\mathbf{\hat r}};\left\{ {{{\mathbf{R}}_{i}}} \right\}} \right) }{{\hat G}^\alpha }\left( E \right){V_{short}}\left( {{\mathbf{\hat r}};\left\{ {{{\mathbf{R}}_{i}}} \right\}} \right) ,
\label{eq:sigma_imp}
\end{align}
where ${V_{short}}$ is the scattering potential under the given impurity distribution, $\left\{ {{{\mathbf{R}}_{i}}} \right\}$, and given by Eq.~\eqref{eq:Vshort}. 
In the real-space representation, it is expressed by
%
\begin{align}
& \Sigma _{imp}^\alpha \left( {{\mathbf{r}},{\mathbf{r'}};E} \right)  = \left\langle {\mathbf{r}} \right|\hat \Sigma _{imp}^\alpha \left( E \right)\left| {{\mathbf{r'}}} \right\rangle 
\approx {e^2}{G^\alpha }\left( {{\mathbf{X}},{\boldsymbol\xi} ;E} \right)  
\nonumber \\
& ~~~~~~\times \sum\limits_{l = 1}^{N_d^{tot}} {{v_{sc}}\left( {{\mathbf{X}} + \frac{\boldsymbol\xi }{2} - {{\mathbf{R}}_l}} \right){v_{sc}}\left( {{\mathbf{X}} - \frac{\boldsymbol\xi }{2} - {{\mathbf{R}}_l}} \right)} 
\label{eq:sigma_imp1}
\end{align}
%
where the Wigner coordinates are defined by ${\mathbf{X}} = \left( {{\mathbf{r}} + {\mathbf{r'}}} \right)/2$ and ${\boldsymbol\xi}  = {\mathbf{r}} - {\mathbf{r'}}$. 
In this expression, we have ignored the correlation among different impurities and only considered the correlation from the same impurity. This (Born) approximation is consistent with the fact that $v_{sc}$ is the short-ranged potential and does not overlap with other impurities. 
Using the Fourier expansion of $v_{sc}$ defined by
\begin{align}
{v_{sc}}\left( {\mathbf{r}} \right)  = \int {\frac{{{d^3}q}}{{{{\left( {2\pi } \right)}^3}}}{{\tilde v}_{sc}}\left( {\mathbf{q}} \right){e^{i{\mathbf{q}} \cdot {\mathbf{r}}}}} ,
\label{eq:vsc_xfrom}
\end{align}
%
and the fact that $v_{sc}$ is a real function, Eq.~\eqref{eq:sigma_imp1} is rewritten as
%
\begin{align}
& \Sigma _{imp}^\alpha \left( {{\mathbf{r}},{\mathbf{r'}};E} \right) = {e^2}\sum\limits_{l = 1}^{N_d^{tot}} {\int {\frac{{{d^3}Q{d^3}p}}{{{{\left( {2\pi } \right)}^6}}}{\kern 1pt} } {e^{i{\mathbf{p}} \cdot \left( {{\mathbf{X}} - {{\mathbf{R}}_l}} \right)}}{e^{i{\mathbf{Q}} \cdot \boldsymbol\xi }}} 
\nonumber \\
&~~~ \times {\kern 1pt} {{\tilde v}_{sc}}\left( {{\mathbf{Q}} + \frac{{\mathbf{p}}}{2}} \right)\tilde v_{sc}^*\left( {{\mathbf{Q}} - \frac{{\mathbf{p}}}{2}} \right){G^\alpha }\left( {{\mathbf{X}},\boldsymbol\xi ;E} \right)
\label{eq:sigma_imp2}
\end{align}
%
where ${\mathbf{p}} = {\mathbf{q}} - {\mathbf{q'}}$ and ${\mathbf{Q}} = \left( {{\mathbf{q}} + {\mathbf{q'}}} \right)/2$.
Since $v_{sc} \left( {\mathbf{r}} \right)$ is the short-ranged and, thus, ${{{\tilde v}_{sc}}\left( {\mathbf{q}} \right)}$ is significant only in the cases of $\left| {\mathbf{q}} \right| > q_c$. 
Hence, we may approximate the term in Eq.~\eqref{eq:sigma_imp2} by
\begin{align}
{{\tilde v}_{sc}}\left( {{\mathbf{Q}} + \frac{{\mathbf{p}}}{2}} \right)\tilde v_{sc}^*\left( {{\mathbf{Q}} - \frac{{\mathbf{p}}}{2}} \right) \approx {\left| {{{\tilde v}_{sc}}\left( {\mathbf{Q}} \right)} \right|^2} .
\label{eq:vsc_xfrom2}
\end{align}
Noting that  
${\mathbf{X}}$ is the `center-of-mass' position of ${\mathbf{r}}$ and ${\mathbf{r'}}$ and, thus, ${G^\alpha }\left( {{\mathbf{X}},{\boldsymbol\xi} ;E} \right)$ varies gradually to ${\mathbf{X}}$. 
As a result, we may also approximate that
${G^\alpha }\left( {{\mathbf{X}},{\boldsymbol\xi} ;E} \right) \approx {G^\alpha }\left( {{{\mathbf{R}}_l},{\boldsymbol\xi} ;E} \right) $ in Eq.~\eqref{eq:sigma_imp2} because the integral over $\mathbf{p}$ is dominated around ${\mathbf{X}} \approx {{\mathbf{R}}_l}$.
%
%
This approximation corresponds to the long-wavelength limit implicitly assumed in the traditional NEGF methods. 
Consequently, we obtain 
\begin{align}
{\Sigma ^\alpha }_{imp}\left( {{\mathbf{r}},{\mathbf{r'}};E} \right) = \sum\limits_{l = 1}^{N_d^{tot}} {\delta \left( {{\mathbf{X}} - {{\mathbf{R}}_l}} \right){C_{\mathbf{X}}}\left( {\boldsymbol\xi}  \right) {G^\alpha }\left( {{\mathbf{X}},{\boldsymbol\xi} ;E} \right)} ,
\label{eq:sigma_impf}
\end{align}
where the correlation function ${C_{\mathbf{X}}}\left( {\boldsymbol\xi}  \right)$ is defined and given by
\begin{align}
{C_{\mathbf{X}}}\left( {\boldsymbol\xi}  \right) & = e^2 \int {\frac{{{d^3}Q}}{{{{\left( {2\pi } \right)}^3}}}{e^{i{\mathbf{Q}} \cdot {\boldsymbol\xi} }}{{\left| {{{\tilde v}_{sc}}\left( {\mathbf{Q}} \right)} \right|}^2}{\kern 1pt} } 
\nonumber \\
& = e^2 \int  {{d^3}r{\kern 1pt} {v_{sc}}\left( {{\mathbf{r}} + \frac{\boldsymbol\xi }{2}} \right){\kern 1pt} {v_{sc}}\left( {{\mathbf{r}} - \frac{\boldsymbol\xi }{2}} \right)} .
\label{eq:CX_xi}
\end{align}
This is a consequence of the Wiener-Khintchine theorem and, interestingly, the second line of Eq.~\eqref{eq:CX_xi} has the same expression of the ensemble average of the (short-range) potential correlation if $v_{sc}$ were a function of random variables~\cite{Reif1965,Reichl1998}. 
%
%
%
Also, since $v_{sc}$ is the potential at position $\mathbf{X}$ and the screening wavenumber $q_c$ included in ${\tilde v}_{sc}$ or $v_{sc}$ depends on $\mathbf{X}$, 
we put the subscript $\mathbf{X}$ in ${C_{\mathbf{X}}}$ to remind that it implicitly depends on the `center-of-mass' position ${\mathbf{X}}$  in the Wigner coordinates.
Under the bulk approximation where $v_{sc}$ is given by the Yukawa potential, ${C_{\mathbf{X}}}$ is analytically expressed by
\begin{align}
{C_{\mathbf{X}}}\left( {\boldsymbol\xi}  \right) = {\left( {\frac{{{e^2}}}{{{\varepsilon _s}}}} \right)^2}\frac{1}{{8\pi }}\frac{1}{{{q_c}}}{e^{ - {q_c}\left| {\boldsymbol\xi}  \right|}}, 
\label{eq:CX_bulk}
\end{align}
where the screening length $q_{c}^{-1}$ is treated by the Debye-H\"{u}ckel and Thomas-Fermi models for nondegenerate and degenerate regimes, respectively~\cite{Sano2024}. 

In short, Eq.~\eqref{eq:sigma_impf} is the self-energy due to impurity scattering under discrete impurities, which accounts for the spatial locality of impurities in a semiconductor substrate. 
It should be noted that the present self-energy is expressed in terms of the Wigner coordinates. As a result, the self-energy given by Eq.~\eqref{eq:sigma_impf} becomes nonlocal in the real-space representation, $\left\{ {\left| {\mathbf{r}} \right\rangle } \right\}$.

\subsection{Self-averaging and reduction to the scattering rate from Fermi's golden rule}
\label{sec:redFermi}
It is now straightforward to show that the proposed model of the self-energy due to discrete impurity scattering  leads, by self-averaging over all possible impurity distributions, to the well-known impurity scattering rate derived from Fermi's golden rule. 

Performing the average over (inhomogeneous) impurity distributions, the averaged self-energy due to impurity scattering  is calculated by 
%
\begin{align}
& \left\langle {{\Sigma ^\alpha }_{imp}\left( {{\mathbf{r}},{\mathbf{r'}};E} \right)} \right\rangle  \hspace*{30mm}
\nonumber \\
& ~~~ = \int  \cdots  \int_\Omega  {\left[ {\prod\limits_{l = 1}^{N_d^{tot}} {{P_d}\left( {{{\mathbf{R}}_l}} \right){d^3}{{\mathbf{R}}_l}} } \right]} {\Sigma ^\alpha }_{imp}\left( {{\mathbf{r}},{\mathbf{r'}};E} \right) 
\nonumber \\
& ~~~ = \bar N_d^ + \left( {\mathbf{X}} \right){C_{\mathbf{X}}}\left( \boldsymbol\xi  \right){G^\alpha }\left( {{\mathbf{X}},{\boldsymbol\xi} ;E} \right) 
\label{eq:sv_sigma}
\end{align}
%
where $\Omega$ is the volume of the device substrate and ${P_d} \left( {\mathbf{R}} \right) $ is the probability density of finding a donor impurity at position ${\mathbf{R}}$ and defined by
\begin{align}
{P_d}\left( {\mathbf{R}} \right) = \frac{{\bar N_d^ + \left( {\mathbf{R}} \right)}}{{N_d^{tot}}}.
\end{align}
Here, $\bar N_d^ + \left( {\mathbf{R}} \right)$ is the `jellium' donor density, namely, the donor density under the long-wavelength limit.
Under the homogeneous impurity configurations, ${P_d}\left( {\mathbf{R}} \right)$ reduces to the well-known expression, ${P_d} = 1/ \Omega$. 
We would like to stress  that, after taking the self-average, ${G^\alpha }\left( {{\mathbf{X}},{\boldsymbol\xi} ;E} \right) $ in Eq.~\eqref{eq:sv_sigma} is replaced with the one under the self-averaged Hartree potential $V_{long}$ which could be very different from the potential obtained by the jellium impurity density $\bar N_{d}^{+} \left( {\mathbf{R}} \right)$ if the doping profiles are strongly inhomogeneous in space~\cite{Sano2024}.

Taking the Fourier-transform with respect to $\boldsymbol\xi$, Eq.~\eqref{eq:sv_sigma} becomes
\begin{align}
&\left\langle {{\Sigma ^\alpha}_{imp}\left( {{\mathbf{X}},{\mathbf{k}};E} \right)} \right\rangle
 \nonumber \\
&~~~~ = \bar N_d^ + \left( {\mathbf{X}} \right)\frac{1}{\Omega }\sum\limits_{\mathbf{q}} {{{\left| {e{{\tilde v}_{sc}}\left( {\mathbf{q}} \right)} \right|}^2}{G^\alpha}\left( {{\mathbf{X}},{\mathbf{k}} - {\mathbf{q}};E} \right)} .
\label{eq:sv_sigma_k}
\end{align}
%
Notice that both $\bar N_d^ + \left( {\mathbf{X}} \right)$ and ${G^\alpha}\left( {{\mathbf{X}},{\mathbf{k}} - {\mathbf{q}};E} \right)$ become dependent on the `center-of-mass' position ${\mathbf{X}}$ and, hence, they are also nonlocal in the real-space representation. 

Suppose that $\alpha =r$ and $G^r$ is replaced by the unperturbed retarded Green's function of the device region, which is given by
${g^r}\left( {{\mathbf{k}};E} \right) = {\left( {E + i{0^ + } - {\varepsilon _{\mathbf{k}}}} \right)^{ - 1}}$ with ${\varepsilon _{\mathbf{k}}} = {\hbar ^2}{{\mathbf{k}}^2}/\left( {2m} \right)$. 
Notice that ${g^r}\left( {{\mathbf{k}};E} \right)$ is the retarded Green's function under the homogeneous structures and independent on $\mathbf{X}$. Hence, 
this replacement corresponds to taking the long-wavelength limit {\em locally}, as stressed in \cite{Sano2018}.
The impurity scattering rate under the self-averaged impurity  is then given by
\begin{align}
\frac{1}{{{\tau _{\mathbf{X}}}\left( {{\varepsilon _{\mathbf{k}}}} \right)}} & =  - \frac{2}{\hbar }\operatorname{Im} \left[ {\left\langle {{\Sigma ^r}_{imp}\left( {{\mathbf{X}},{\mathbf{k}};{\varepsilon _{\mathbf{k}}}} \right)} \right\rangle } \right]
\nonumber \\
&  = \frac{{2\pi }}{\hbar }\bar N_d^ + \left( {\mathbf{X}} \right)\frac{1}{\Omega }\sum\limits_{\mathbf{q}} {{{\left| {e{{\tilde v}_s}\left( {{\mathbf{k}} - {\mathbf{q}}} \right)} \right|}^2}} \delta \left( {{\varepsilon _{\mathbf{k}}} - {\varepsilon _{{\mathbf{k}} - {\mathbf{q}}}}} \right) .
\label{eq:Fermi}
\end{align}
This is {\em nearly} identical to the expression of the impurity scattering rate due to the screened potential derived from Fermi's golden rule in the literature~\cite{Jacoboni2010book,Fischetti2016book}.  

There is, however, a notable difference between Eq.~\eqref{eq:Fermi} and the expression in the literature. 
In the latter case, the position dependence of the scattering rate is given by $\bar N_d^ + \left( {\mathbf{r}} \right)$, rather than $\bar N_d^ + \left( {\mathbf{X}} \right)$, and, thus, the scattering rate becomes local in space. 
Furthermore, this position dependence found in the literature is introduced in an {\it ad hoc} empirical basis. 
To the author's knowledge, this is the first time to systematically derive the (global) position-dependence of the impurity scattering rates under {\em nonuniform} impurity profiles and to clarify its physical meaning. 
We would like to stress again that 
the self-energies given by Eq.~\eqref{eq:sv_sigma} are also nonlocal {\em even after self-averaging}. 
As a consequence, the local approximation of the self-energy in the real-space representation due to either self-averaged or non self-averaged impurity tends to underestimate the dephasing effects 
and leads to overestimation of the mobility due to impurity scattering, as we shall show in Sec.~\ref{sec:R&D}.

\section{Quasi-1D approximation of NEGF}
\label{sec:quasi-1D}
We assume that the device region has a thin cylindrical wire structure throughout the present numerical calculations. 
Due to random distributions of discrete impurities in a substrate, the potential profiles inside the wires are inherently three-dimensional (3-D) and, thus, the 3-D NEGF simulation would be mandatory to predict the transport properties quantitatively. 
However, such calculations require a heavy computational burden because of asymmetric potential profiles which prevent from using any symmetric properties in numerical calculations. Fortunately, the variations in the electrostatic potential and the effective donor density over the cross-sectional area are gradual. Thus, the potential profiles along the different axes in the cross-sectional area 
are similar to each other, owing to the small radius of cylindrical wires. 
In the present NEGF simulations, therefore, the quasi-one-dimensional (quasi-1D) approximation is employed to reduce a computational burden: The electrostatic potential along the cylindrical axis from the 3-D Poisson equation is used as a frozen potential. 
Since the quasi-1D electron density found from the NEGF simulation cannot be an input to the 3-D Poisson equation, the self-consistent loop between the `long-range' Poisson equation and the NEGF is removed in the present simulations. This approximation may be justified as far as the electron transport under the linear response regime is concerned. 

Since the quantum wire is connected with semi-infinite leads at the left and right ends of the wire,  electrons with the Fermi-Dirac distributions of the chemical potentials $\mu_{L}$ and $\mu_{R}$ are injected from the leads into the wire (device) region. 
Electrons then propagate only in the wire direction, and the (long-range) potential fluctuations in the transverse direction are ignored.  
This is, however, not so critical 
because such potential fluctuations 
do not directly affect the phase coherence in the propagating direction~\cite{sano2017variability}.
Imposing the boundary condition such that the electron wave function vanishes on the cylindrical surface, we expand the basis vector in the real-space representation in terms of the Bessel series as
\begin{align}
\left| {\mathbf{r}} \right\rangle  & = \left| z \right\rangle  \otimes \sum\limits_{l,n} {\left| {l,n} \right\rangle \left\langle {l,n} \right|\left. {\boldsymbol\rho}   \right\rangle }  \simeq \left| z \right\rangle  \otimes \left| {0,1} \right\rangle \left\langle {0,1} \right|\left. {\boldsymbol\rho}  \right\rangle  
 \nonumber \\
& \equiv \left| {z,0,1} \right\rangle {\xi _{01}}^*\left( {\boldsymbol\rho}  \right) ,
\label{eq:FBseries}
\end{align}
where ${\boldsymbol\rho} = \left( {\rho} ,{\phi} \right) $ is a radial vector in the cylindrical coordinates and $\xi _{ln}$ is 
the subband wave function expressed by
\begin{align}
{\xi _{ln}}\left( {\boldsymbol\rho}  \right) = \left\langle {\boldsymbol\rho}  \right|\left. {l,n} \right\rangle  = \frac{1}{{\sqrt \pi  {\rho _s}{J_{l + 1}}\left( {{x_{ln}}} \right)}}{J_l}\left( {{x_{ln}}\frac{\rho }{{{\rho _s}}}} \right) e^{i l \phi}.
\label{eq:subwf}
\end{align}
Here, $\rho _s$ is the radius of the wire, $J_l$ is the $l $-th order Bessel function, and $x_{ln}$ is the $n$-the zero of $J_l$. Under the quasi-1D approximation, the cylindrical wire becomes symmetric about the wire axis and, thus, Eq.~\eqref{eq:FBseries} is approximated by the lowest subband, $l=0$ and $n=1$. 

\begin{figure}[tb]%
 \begin{minipage}{1\hsize}
 \begin{center}
  \includegraphics[width=6.8cm]{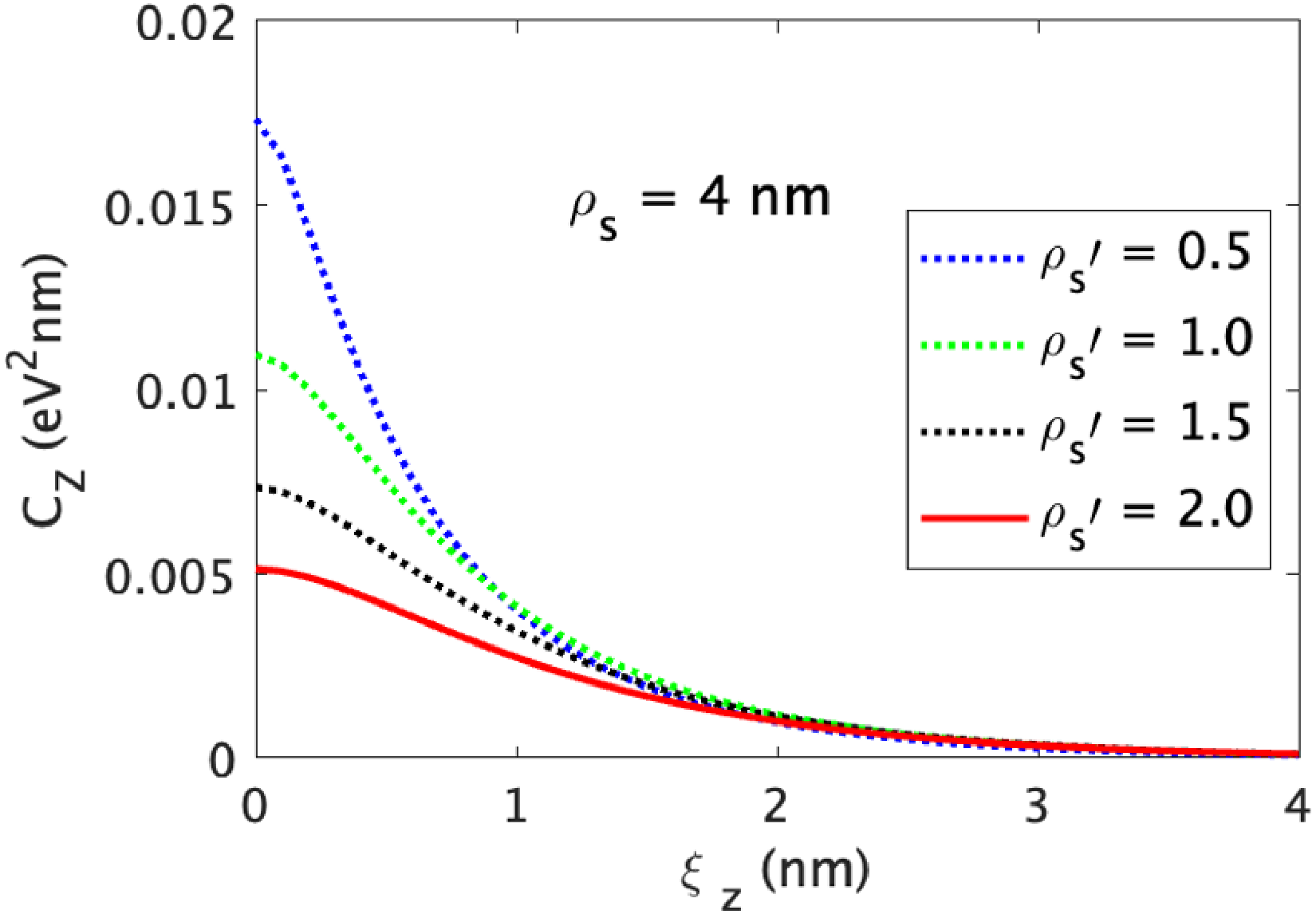}
  \caption{
 Quasi-1D correlation function as a function of the relative distance $\xi _z$. The correlation functions for $\rho _s ' = 0.5, 1, 1.5, 2$ are shown. 
} 
  \label{fig:cz}
 \end{center}
 \end{minipage}
 \end{figure}

The effective scattering potential $v_{sc}^{\rm eff}$ under the quasi-1D approximation due to a single-impurity at the origin is then given by
%
\begin{align}
& \left\langle {z,0,1} \right|{v_{sc}}\left( {{\mathbf{\hat r}}} \right)\left| {z',0,1} \right\rangle  = \frac{e}{{{\varepsilon _s}}}\frac{1}{{{\rho _s}^2}}\frac{1}{{2\pi {{\left\{ {{J_1}\left( {{x_{01}}} \right)} \right\}}^2}}}
\nonumber \\
& \hspace*{10mm} \times \int_0^{{\rho _s}} {d\rho \rho \frac{{{e^{ - {q_c}\sqrt {{\rho ^2} + {z^2}} }}}}{{\sqrt {{\rho ^2} + {z^2}} }}{{\left[ {{J_0}\left( {{x_{01}}\frac{\rho }{{{\rho _s}}}} \right)} \right]}^2}} \delta \left( {z - z'} \right)
\nonumber \\
& ~~~ \equiv {v_{sc}^{\rm eff}}\left( z \right)\delta \left( {z - z'} \right) .
\label{eq:eff_vsc}
\end{align}
%
Since $v_{sc}^{\rm eff}$ is dependent only on $z$, the correlation function defined by Eq.~\eqref{eq:CX_xi} becomes
\begin{align}
{C_{\mathbf{X}}}\left( {\boldsymbol\xi }  \right) & \simeq {e^2}\int {{d^3}rv_{sc}^{\rm eff}\left( {z + \frac{{{\xi _z}}}{2}} \right)\,v_{sc}^{\rm eff}\left( {z - \frac{{{\xi _z}}}{2}} \right)}  
\nonumber \\ 
& = S_c {\left( {\frac{{{e^2}}}{{{\varepsilon _s}}}} \right)^2}\frac{1}{{{\rho _s}}}\frac{1}{{4{\pi ^2}{{\left\{ {{J_1}\left( {{x_{01}}} \right)} \right\}}^4}}}
\nonumber \\ 
& ~~ \times \int_{ - \infty }^\infty  {d\zeta {I_v}\left( {\zeta  + \frac{{{q_c}{\xi _z}}}{2}} \right){I_v}\left( {\zeta  - \frac{{{q_c}{\xi _z}}}{2}} \right)} 
\nonumber \\ 
& \equiv S_c {C_Z}\left( {{\xi _z}} \right) ,
\label{eq:1DdC}
\end{align}
where ${C_Z}\left( {{\xi _z}} \right)$ is the quasi-1D correlation function and the cross-sectional area of the wire is given by ${S_c} = \pi {\rho _s}^2$.
$I_v$ in Eq.~\eqref{eq:1DdC} is defined by  
\begin{align}
{I_v}\left( \zeta  \right) = \frac{1}{{{{\left( {{\rho _s}^\prime } \right)}^{3/2}}}}\int_0^{{\rho _s}^\prime } {d\rho '\rho '\frac{{{e^{ - \sqrt {{{\rho '}^2} + {\zeta ^2}} }}}}{{\sqrt {{{\rho '}^2} + {\zeta ^2}} }}{{\left[ {{J_0}\left( {{x_{01}}\frac{{\rho '}}{{{\rho _s}^\prime }}} \right)} \right]}^2}} ,
\label{eq:Iv}
\end{align}
where ${\rho _s}^\prime  = {q_c}{\rho _s}$ and $\zeta  = {q_c}z $. 
The quasi-1D correlation functions are shown in Fig.~\ref{fig:cz} 
for ${\rho _s}' = q_{c}\rho _s = 0.5, 1, 1.5, 2$. 
In the present case, the screening length is smaller than the cylindrical radius ($\rho _s$ = 4 nm) and, thus, throughout the present NEGF simulations, we assume that ${\rho _s}^\prime = 2$. 

Under the quasi-1D approximation, the transverse coordinates of impurity's position in Eq.~\eqref{eq:sigma_impf} must be averaged over the cross-sectional area of the cylindrical wire. After averaging, we obtain 
\begin{align}
& \int \cdots \int _{S_c} {\left[ {\prod\limits_{m = 1}^{N_d^{tot}} {\frac{{{d^2}{R_{m,\rho }}}}{{S_c}}} } \right]} \sum\limits_{l = 1}^{N_d^{tot}} {\delta \left( {{\mathbf{X}} - {{\mathbf{R}}_l}} \right){C_{\mathbf{X}}}\left( {\boldsymbol\xi}  \right)}~~~
\nonumber \\
& \hspace*{15mm} = \sum\limits_{l = 1}^{N_d^{tot}} {\delta \left( {Z - {R_{l,z}}} \right){C_Z}\left( {{\xi _z}} \right)} ,
\label{eq:1DdCr}
\end{align}
where $Z$, $R_{l,z}$ and $\xi _z$ are the $z$-components of ${\mathbf{X}}$, ${\mathbf{R}}_l$, and ${\boldsymbol\xi} $, respectively. $\mathbf{R}_{m,\rho }$ is the position vector of the $m$-th impurity in the radial direction of the cylindrical coordinates.
After all, the quasi-1D self-energy due to impurity scattering is given by 
\begin{align}
& \Sigma _{imp}^\alpha \left( {z,z';E} \right) \equiv \left\langle {z,0,1} \right|\hat \Sigma _{imp}^\alpha \left( E \right)\left| {z',0,1} \right\rangle  
\nonumber \\
& ~~~~~ = \sum\limits_{l = 1}^{N_d^{tot}} {\delta \left( {Z - {R_{l,z}}} \right){C_Z}\left( {{\xi _z}} \right){G^\alpha }\left( {z,z';E} \right)} ,
\label{eq:1Dsigimp}
\end{align}
with $Z = \left( {z + z'} \right)/2$ and ${\xi _z} = z - z'$. 
{
Notice that the nonlocality in the self-energy comes from the ${\xi _z}$ dependence of the correlation function.}
Here, the quasi-1D Green's function is defined by 
\begin{align}
{G^\alpha }\left( {z,z';E} \right) \equiv \left\langle {z,0,1} \right|{{\hat G}^\alpha }\left( E \right)\left| {z',0,1} \right\rangle  .
\label{eq:1DG}
\end{align}

Furthermore, the left and right leads are semi-infinite 1-D wires and the quasi-1D self-energy due to the L(R) contact with the lead is then given by
\begin{align}
\Sigma _{L(R)}^\alpha \left( {z,z';E} \right) \equiv \left\langle {z,0,1} \right|\hat \Sigma _{L(R)}^\alpha \left( E \right)\left| {z',0,1} \right\rangle  .
\label{eq:1DSigmaLR}
\end{align}
Specifically, under the 1-D lattice system, the retarded self-energy is nonzero only if $z = z' = 0$ or $L_c$ and it is given by 
\begin{align}
\Sigma _{L(R)}^r \left( {0 \,(L_c),0 \,(L_c);E} \right)  = -\frac{t_0}{a}  {e^{ika}} ,
\label{eq:1DSigmaLR^r}
\end{align}
where $t_0 = \hbar ^2 / \left( 2 m a^2  \right)  $, 
$a$ is the lattice spacing, and $k$ is the wave-number of electrons in the tight-binding approximation, which satisfies the following energy dispersion;
\begin{align}
\cos  \left( ka \right) = 1-\frac{E-{\varepsilon _{01}} + eV_{long}}{2 t_0} .
\label{eq:1DEdip}
\end{align}
Here, $\varepsilon _{01}$ is the lowest subband energy given by 
${\varepsilon _{01}} = {\hbar ^2}{x_{01}}^2/\left( {2m{\rho _s}^2} \right)$ 
and $V_{long}$ is evaluated at $z=0$ ($L_c$) for the left (right) contact. 
Similarly, the lesser self-energy is given by
\begin{align}
 \Sigma _{L(R)}^< \left( {0 \,(L_c),0 \,(L_c);E} \right) = i f_{FD}^{L(R)}\left( E \right)  
\frac{2 t_0}{a} \sin \left( ka  \right) .
\label{eq:1DSigmaLR^<}
\end{align}
Notice that both Eqs.~\eqref{eq:1DSigmaLR^r} and \eqref{eq:1DSigmaLR^<} are directly dependent on the lattice spacing $a$ because of the usage of the lattice system.

Consequently, the quasi-1D retarded Green's function is given by the usual Dyson equation; 
\begin{align}
& {G^r}\left( {z,z';E} \right) = G_0^r\left( {z,z';E} \right) + \int _{L_c} {d{z_1}d{z_2}G_0^r\left( {z,{z_1};E} \right)} 
\nonumber \\
&  ~~ \times \left[ {\Sigma _{L/R}^r\left( {{z_1},{z_2};E} \right) + \Sigma _{imp}^r\left( {{z_1},{z_2};E} \right)} \right]{G^r}\left( {{z_2},z';E} \right),
\label{eq:1DGr}
\end{align}
where $G_0^r\left( {z,z';E} \right) $ is the unperturbed retarded Green's function of the device region and given by
\begin{align}
{G^r_0} \left( {z, z';E} \right) & \equiv \left\langle {z,0,1} \right|\frac{1}{{E + i{0^ + } - {{\hat H}_D} }}\left| {z',0,1} \right\rangle 
\nonumber \\
& =  \left\langle {z} \right|\frac{1}{{E  + i{0^ + } - {{\hat h}_D} -{\varepsilon _{01}}}}\left| {z'} \right\rangle .
\label{eq:1DGr0}
\end{align}
Here, $\hat h _D$ is the 1-D Hamiltonian of the device region and given by 
\begin{align}
{{\hat h}_D} = \frac{{{\hat p}_z}^2}{2m} 
- e{V_{long}}\left( {\hat z};{\left\{ R_{l,z} \right\}}  \right) .
\label{eq:1DhD}
\end{align}
Notice that ${V_{long}}\left(z; {\left\{ R_{l,z} \right\}} \right) $ is the `long-range' electrostatic potential {\em along the wire axis} under discrete impurities. 
We would like to stress that ${G^r_0}$ is different from the usual unperturbed retarded Green's function under `jellium' impurity; the `long-range' part of the impurity potential is explicitly included in ${G^r_0}$ 
in an unperturbed way.
Similarly, the quasi-1D lesser Green's function is derived from Eq.~\eqref{eq:G<} and given by the Keldysh equation;
%
\begin{align}
& {G^ < }\left( {z,z';E} \right) = \int _{L_c} {d{z_1}d{z_2}{G^r}\left( {z,{z_1};E} \right)} 
\nonumber \\
& ~~ \times 
\left[ {\Sigma _{L/R}^ < \left( {{z_1},{z_2};E} \right) + 
\Sigma _{imp}^<\left( {{z_1},{z_2};E} \right)} \right] {G^a}\left( {{z_2},z';E} \right) .
\label{eq:1DG<}
\end{align}
%
%
The present theoretical framework of the quasi-1D NEGF simulations is based on Eqs~\eqref{eq:1DGr} and \eqref{eq:1DG<} along with Eqs.~\eqref{eq:1Dsigimp}, \eqref{eq:1DSigmaLR^r}, and \eqref{eq:1DSigmaLR^<}. 
Both equations of ${G^ r }$ and ${G^ < }$ are nonlinear and, thus, the self-consistent calculations are required. 

The local density of states (LDOS) at position $z$ is obtained from 
\begin{align}
{D} \left( {z,E} \right)  = - \frac{1}{\pi } 
\operatorname{Im}  \left[  G^r  \left( {z,z ;E} \right) \right] 
\label{eq:ldos}
\end{align}
and the electron density  {\em per unit volume} is calculated by 
\begin{align}
n\left( z \right) & = \frac{2}{S_c}\int {{d^2}{\boldsymbol\rho} } \left[ { - i\int {\frac{{dE}}{{2\pi }}{G^ < }\left( {{\mathbf{r}},{\mathbf{r}};E} \right)} } \right] 
\nonumber \\
& = \frac{2}{S_c}\left[ { - i\int {\frac{{dE}}{{2\pi }}{G^ < }\left( {z,z;E} \right)} } \right] ,
\label{eq:n_3d}
\end{align}
where the factor 2  in Eq.~\eqref{eq:n_3d} stands for the spin degeneracy of electrons. 
Notice that the electron density above  is the one averaged over the cross-sectional area of the wire. 
%
The electric current at the left (right) end of the wire is calculated by the usual expression~\cite{Haug2008,Datta2005};
\begin{align}
{I_{L(R)}} = \frac{e}{{\pi \hbar }}\int {dE} \,{\text{Tr}}\left[ {{G^ > }\Sigma _{L(R)}^ <  - {G^ < }\Sigma _{L(R)}^ > } \right]  .
\label{eq:IcLR}
\end{align}
The total current is then evaluated by $I = \left( {{I_L} - {I_R}} \right)/2$. 
The electron mobility under the given impurity distribution is estimated by
\begin{align}
\mu  = \frac{{{L_c}}}{{e n\left( 0 \right)\pi {\rho _s}^2}}\frac{I}{{{V_{app}}}} ,
\label{eq:mob}
\end{align}
where $n\left( 0 \right) $ is the electron density at the the left end of the wire and calculated from the self-consistent $G^<$ by Eq.~\eqref{eq:n_3d}.

\section{Results and Discussion}
\label{sec:R&D}
\subsection{Long-range potential fluctuations due to discrete impurities}
\label{sec:epot}
We now apply the present framework to the quantum wire structures with the radius of $\rho _s$ = 4 nm and the length of $L_c =$ 30 nm. 
{
The material parameters are taken from those of Si, namely, $m = 0.32 m_{0}$ and $\varepsilon _s = 11.8 \varepsilon _0$, where $m_0$ and $\varepsilon _0$ are the electron bare mass and the permittivity in vacuum, respectively.}
The wire is doped with donor impurities with the presumed `jellium' density $\bar N_d^+ $,  
and discrete impurities are randomly distributed over the substrate consistently with $\bar N_d^+ $. 
The Poisson equation given by Eq.~\eqref{eq:Poissonlong} for the `long-range' part of the electrostatic potential is solved under equilibrium. 
The effective donor density $N_{d,long}^ +$ is approximated by Eq.~\eqref{eq:Nd_long}. The potential on the cylindrical surface is assumed to be grounded. 
Assuming the linear response regime, a uniform electric field is imposed on the electrostatic potential along the wire-axis direction. 

\begin{figure}[tb]%
 \begin{minipage}{1\hsize}
 \begin{center}
  \includegraphics[width=8.5cm]{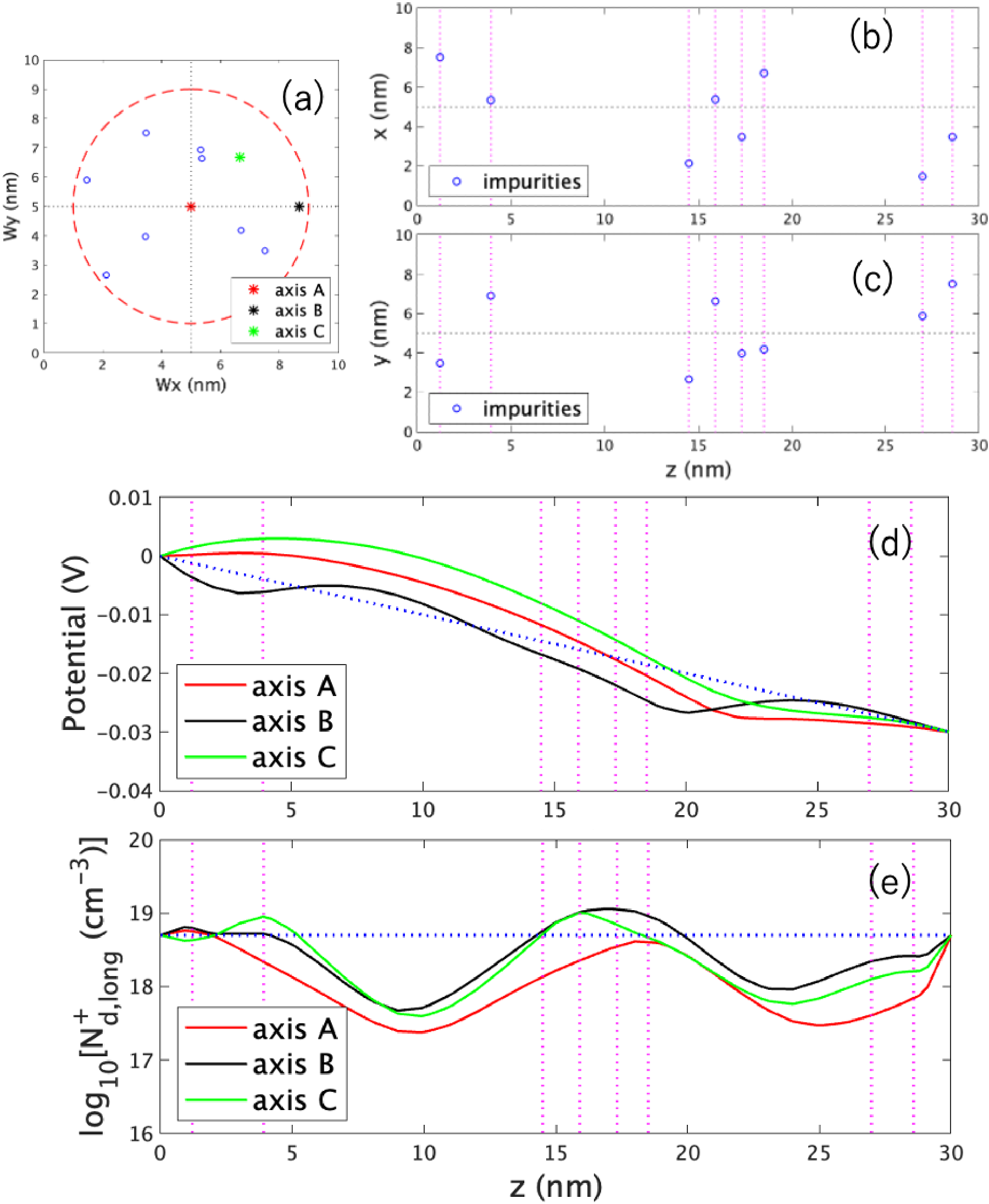}
  \caption{
 Impurity distributions (small empty circles) inside the cylindrical wire with $\rho_{s}=$ 4 nm under $\bar N_d^ + = 5 \times 10^{18}$ cm$^{-3}$. There are 8 impurities in the substrate, which are projected onto the cross-sectional area of the wire in (a), and onto the $z$-$x$ and $y$-$z$ planes in (b) and (c), respectively.  
 The red, black, and green symbols in (a) represent the position of the axis along which the electrostatic potentials with respect to electrons ($-V_{long}$) and the effective donor densities ($N_{d,long}^+$) are shown in (d) and (e). The vertical red dotted lines show the $z$ coordinates of the impurities' positions. 
} 
  \label{fig:impcon}
 \end{center}
 \end{minipage}
 \end{figure}

Figure~\ref{fig:impcon} shows the spatial distributions of impurities inside the quantum wire, the electrostatic potentials, and the effective donor densities under the given impurity distributions. We assume the presumed donor density of $\bar N_d^+ =  5 \times 10^{18} $ cm$^{-3}$, which corresponds to the mean number of impurities inside the device region is 7.5, and the applied voltage of $V_{app} = 30$ mV.
The $z$-axis is taken along the cylindrical wire axis. There are 8 impurities in the substrate and the positions of impurities under this impurity configuration are represented by open circles in (a), which shows the projection of impurity's position onto the cross-sectional area of the wire, and (b) and (c) show those onto the $z$-$x$ and $y$-$z$ planes, respectively. The red, black, and green stars in (a) represent the positions of the axes (denoted as A, B, and C) along which the `long-range' electrostatic potentials with respect to electrons ($-V_{long}$) and the effective donor densities ($N_{d,long}^+$) are shown in (d) and (e). The red vertical dotted lines in (b) to (e) show the positions of impurities.
 
Since the substrate is uniformly doped, the donor density and potential are {\em macroscopically} uniform. 
However, because of the discrete nature of impurities, potential fluctuations associated with the localized impurity potential are observed. It should be noted, however, that the magnitude of the potential fluctuations is the order of a few tens mV (under this particular ${\bar N}_d^+$), owing to the extraction of the `long-range' components of the impurity potential. Therefore, free electrons would easily overcome such small potential barriers and result in the conductivity of the quantum wires at room temperature. 
The {effective} donor densities under the given impurity distribution along the same directions are shown in (e). The donor density spreads over the substrate due the extraction of the `long-range' components from impurity potential, although impurities are localized as point charges. 
We would like to point out that the donor density shown in (e) should be interpreted as the one employed for, say, device simulations where the Poisson equation is always coupled with the transport equation of carriers and the potential thus obtained is the `long-range' self-consistent Hartree potential. 

\begin{figure}[tb]%
 \begin{minipage}{1\hsize}
 \begin{center}
  \includegraphics[width=8.5cm]{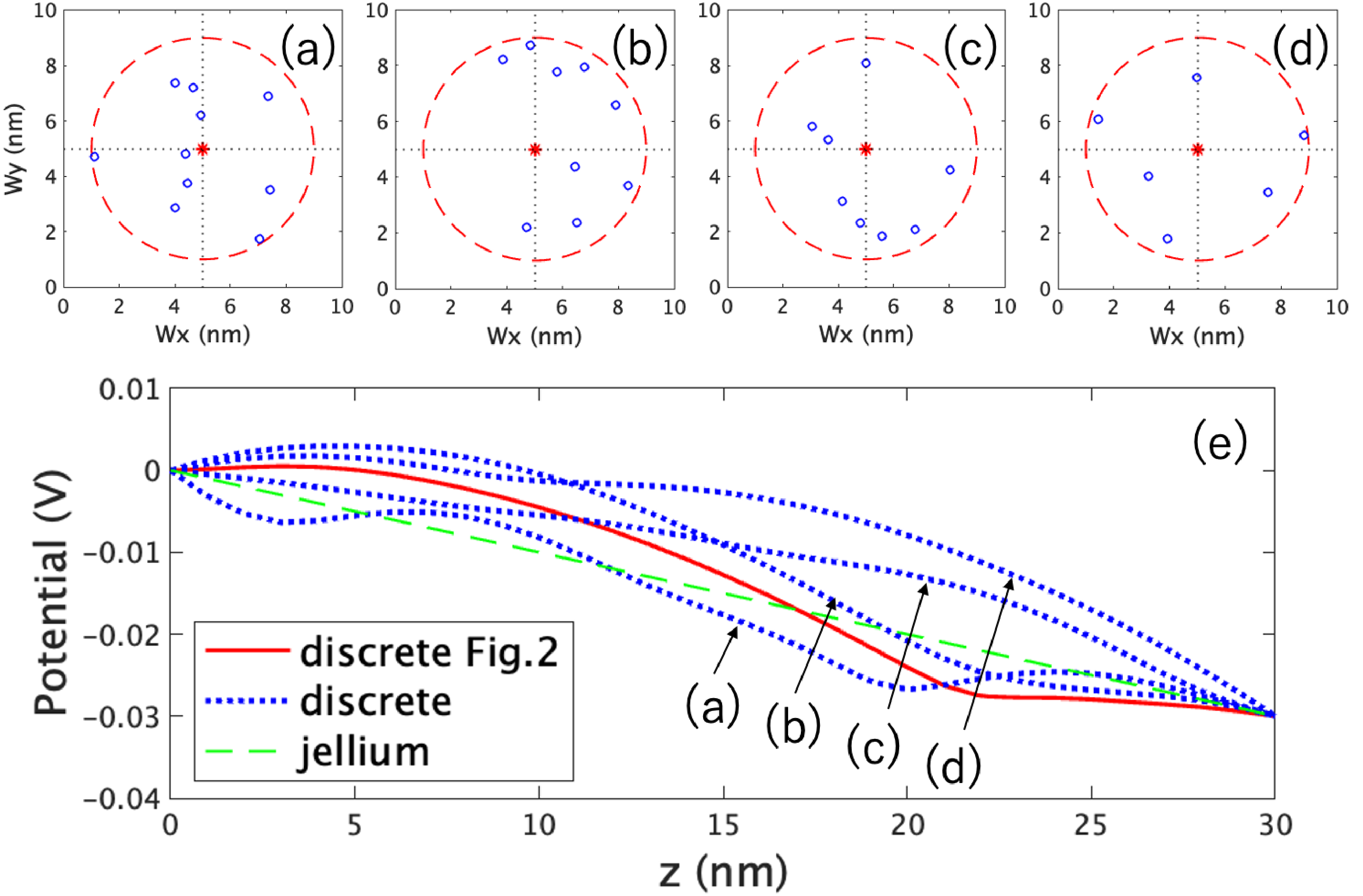}
  \caption{
  (a)-(d) Impurity distributions inside the cylindrical wire with $\rho_{s}=$ 4 nm under $\bar N_d^ + = 5 \times 10^{18}$ cm$^{-3}$. 
(e) Electrostatic potentials with respect to electrons along the wire axis (axis A) under the above impurity distributions. The (red) solid curve represents the potential shown in Fig.~\ref{fig:impcon}. The (green) dashed line denoted as `jellium' represents the uniform potential drop at $V_{app} = 30$ mV.
} 
  \label{fig:impot}
 \end{center}
 \end{minipage}
 \end{figure}

Four different spatial distributions of impurities generated at random under $\bar N_d^ + = 5 \times 10^{18}$ cm$^{-3}$ are shown in Fig.~\ref{fig:impot} (a) to (d), and four (blue) dotted curves in (e) show the electrostatic potentials (with respect to electrons) along the wire axis (axis A) under the given impurity distributions shown in (a) to (d). 
The magnitude of the potential fluctuations is indeed similar to those in Fig.~\ref{fig:impcon} (d). 
\begin{figure}[tb]%
 \begin{minipage}{1\hsize}
 \begin{center}
  \includegraphics[width=8.5cm]{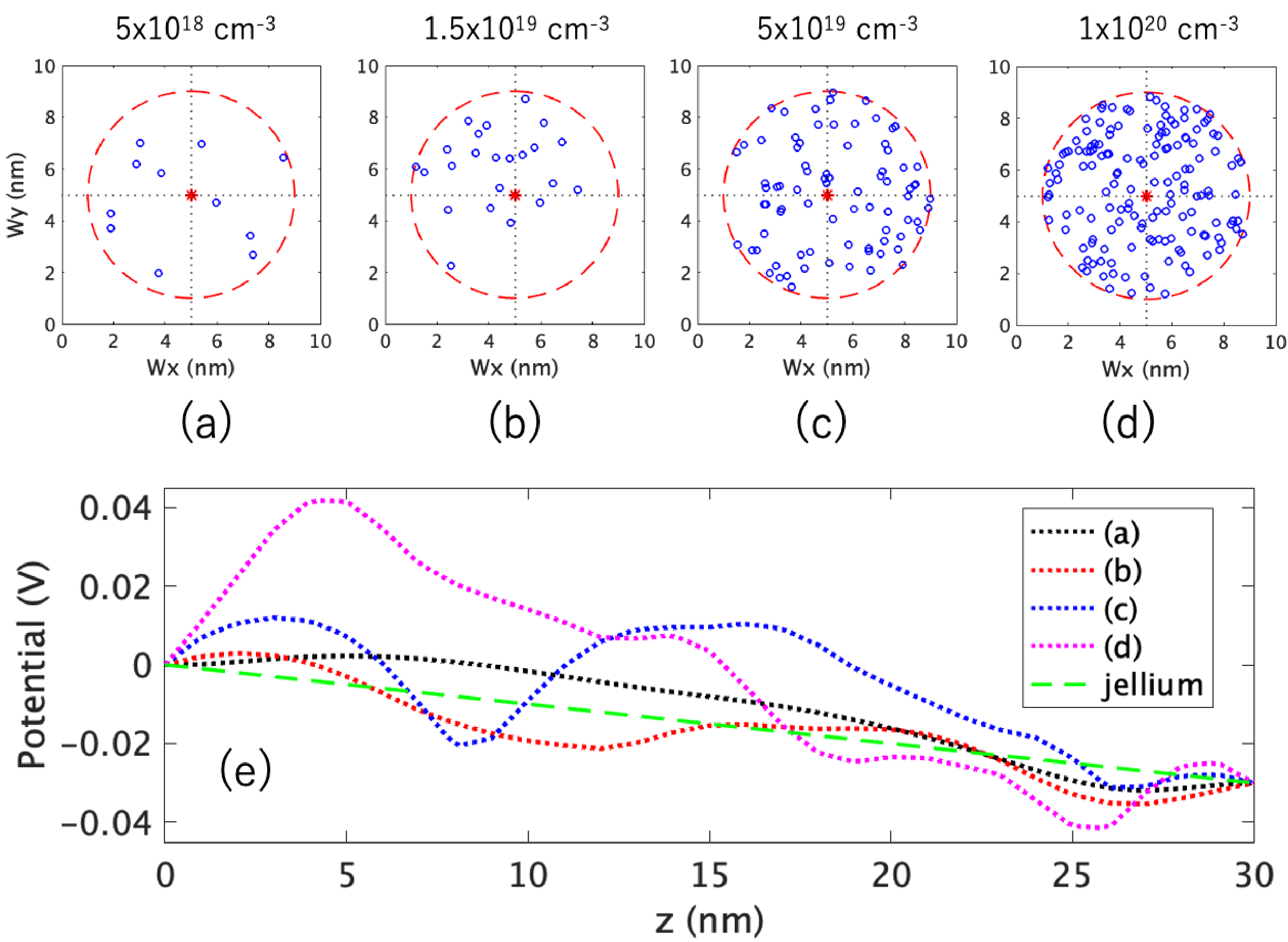}
  \caption{
  (a)-(d) Impurity distributions inside the cylindrical wire with $\rho_{s}=$ 4 nm under $\bar N_d^ + = 5 \times 10^{18},~ 1.5 \times 10^{19},~ 5 \times 10^{19},~ 10^{20}, $ cm$^{-3}$. 
(e) Electrostatic potential with respect to electrons along the wire axis under the impurity densities in (a) to (d).
} 
  \label{fig:npot}
 \end{center}
 \end{minipage}
 \end{figure}%
Since the potential variations are closely correlated with the screening length, the magnitude of the potential fluctuations is strongly dependent on $\bar N_d^ +$. 

Figure~\ref{fig:npot} shows the electrostatic potentials (with respect to electrons) with four different impurity densities, $\bar N_d^ + = 5 \times 10^{18},~1.5 \times 10^{19},~5 \times 10^{19},~10^{20}$ cm$^{-3}$. The potential fluctuations could reach at most one hundred mV for $\bar N_d^ + = 10^{20}$ cm$^{-3}$. Such large potential fluctuations may lead to localizations of electrons in space {\em even at room temperature}.

\subsection{NEGF simulation results and discussion}
\label{sec:NEGF_RD}
The quasi-1D approximation of the NEGF explained in Sec.~\ref{sec:quasi-1D} is applied to cylindrical quantum wire structures with $\rho _{s}= 4$ nm and $L_{c}=30$ nm. The applied voltage is assumed to be $V_{app} = 30$ mV in the present section. 
As explained in Sec.~\ref{sec:epot}, the electrostatic potential along the cylindrical axis found from the 3-D Poisson equation is employed as a fixed potential. 

\begin{figure}[tb]%
 \begin{minipage}{1\hsize}
 \begin{center}
  \includegraphics[width=8cm]{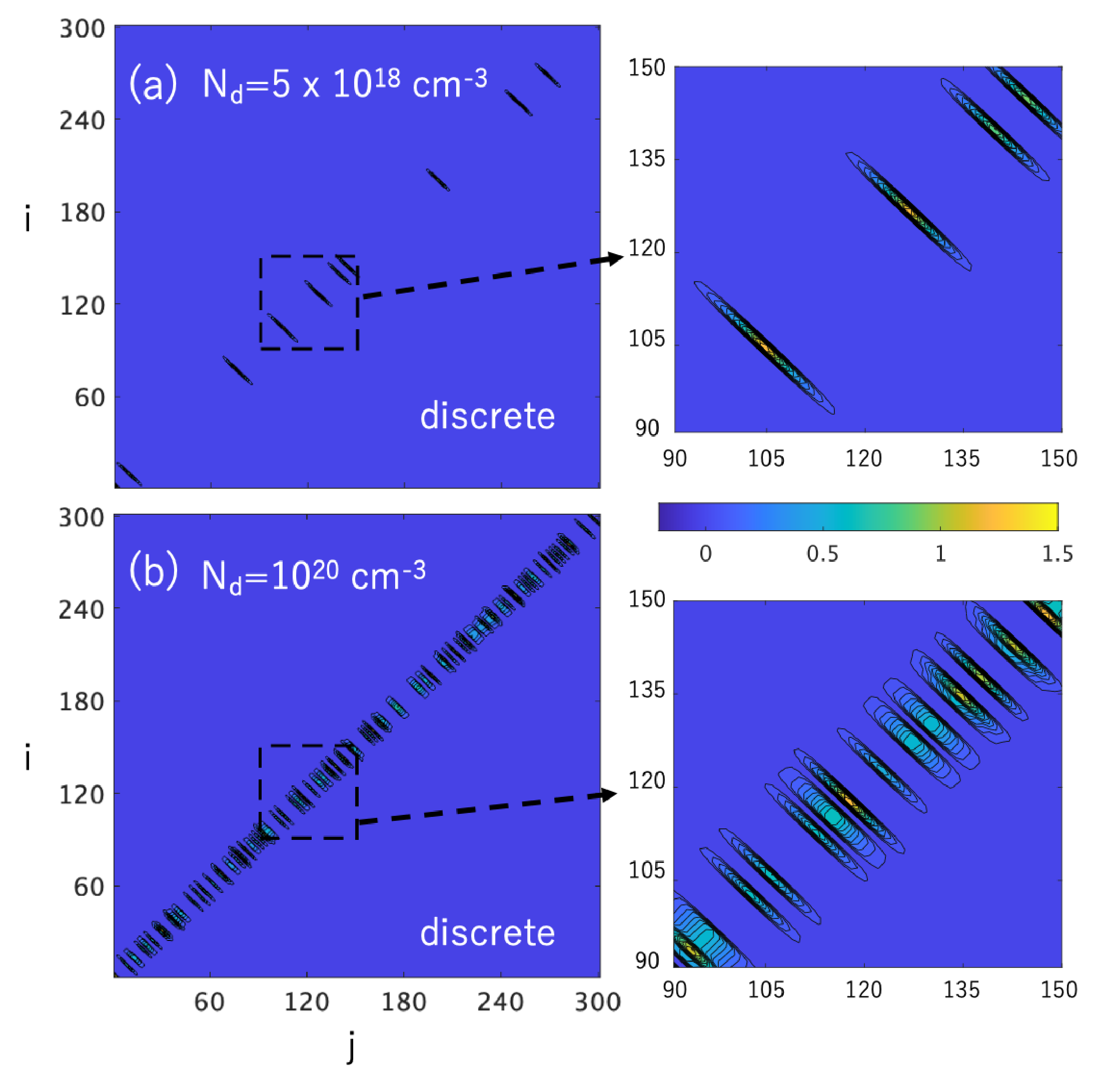}
  \caption{
  Contour plots of the scattering-rate matrices at $E - {\varepsilon _{01}} = 0.05$ eV
  for (a) $\bar N_d^+ = 5 \times 10^{18}$ and (b) $10^{20}$ cm$^{-3}$. 
  The figures on the right are the magnified plots of the region enclosed by dashed lines in the left figures. 
} 
  \label{fig:scmatrix}
 \end{center}
 \end{minipage}
 \end{figure}

The device region is numerically discretized into $N_p$ meshes and, thus, the size of the scattering-rate matrix becomes $ (N_p+1) \times (N_p+1)$.
The element of the scattering-rate matrix due to impurity scattering at electron's energy $E$ is defined by  
\begin{align}
{\Gamma _{imp}}\left( {i,j;E} \right) = - 2 
\operatorname{Im}  \left[ \Sigma _{imp}^r\left( {i a, j a ;E} \right) \right] ,
\label{eq:scmatrx}
\end{align}
where $i, j = 0, 1, \dots, N_p$. Throughout the present simulations, $N_p = 300$ is used. 
Figure~\ref{fig:scmatrix} shows the contour plots of the scattering-rate matrices at $E - {\varepsilon _{01}} = 0.05$ eV under $\bar N_d^+ = 5 \times 10^{18}$ and $10^{20}$ cm$^{-3}$. 
The blue area represents the region whose elements 
are very close to zero. On the other hand, the light blue and yellow areas represent where the elements have large values.  
We would like to stress that the elements of the scattering-rate matrices (and, thus, the self-energy due to impurity scattering) are localized, corresponding to the positions of impurities, as we have intended in the present theory. Furthermore, the matrices are symmetric and non-diagonal, and the values of the off-diagonal elements could be even negative (dark blue regions) although their magnitude is minimal. This kind of structure of the scattering-rate matrices is very similar to the characteristics of the Wigner function which could be negative where the phase interference is strong~\cite{Sano2021pre,Frensley1990}. Such characteristics are purely quantum-mechanical and in sharp contrast with the conventional treatment of impurity scattering, in which the scattering-rate matrix is diagonal, positive definite, and delocalized over space due to self-averaging over impurity distributions.

\begin{figure}[tb]%
 \begin{minipage}{1\hsize}
 \begin{center}
  \includegraphics[width=7.5cm]{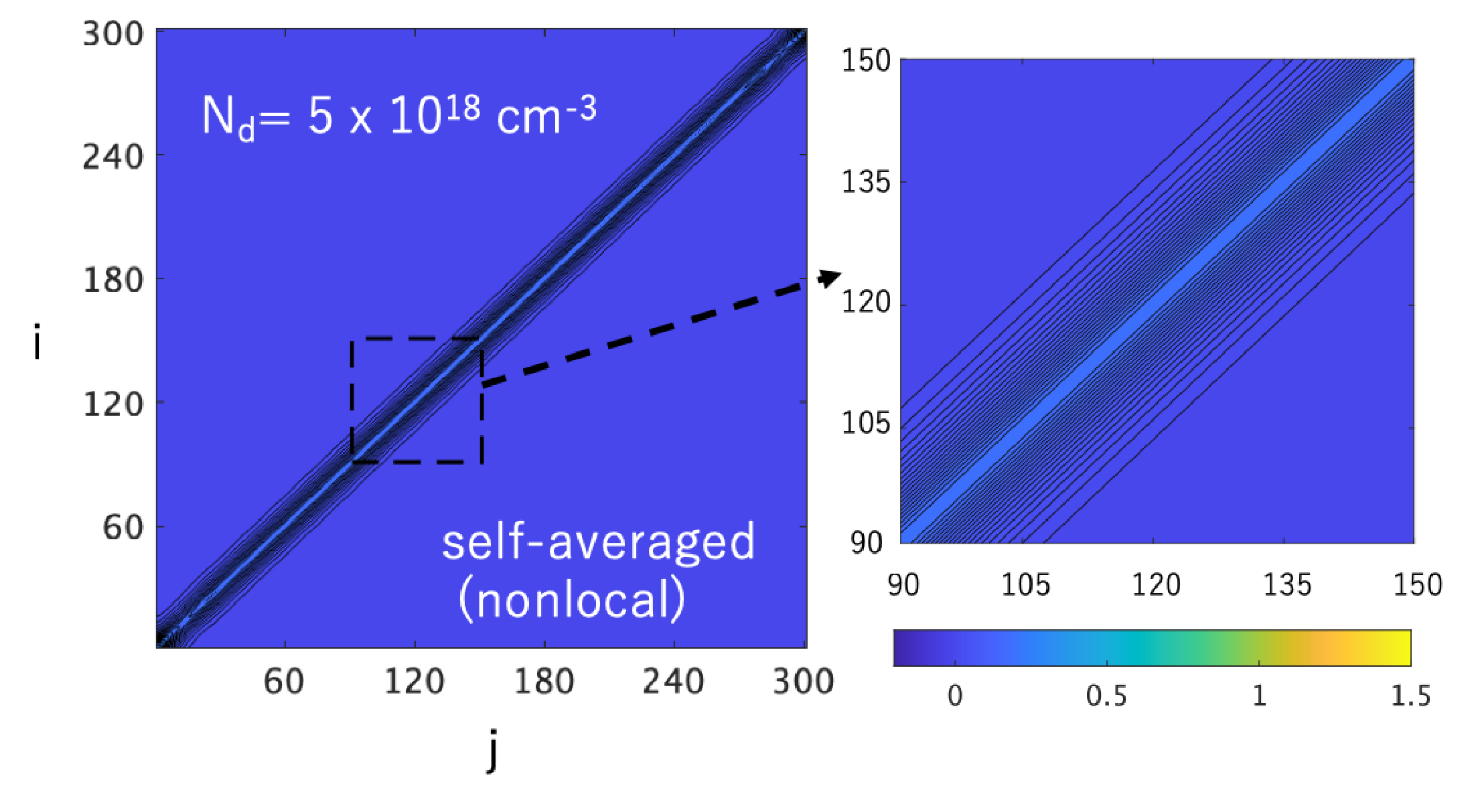}
  \caption{
  Contour plots of the scattering-rate matrix at $E - {\varepsilon _{01}} = 0.05$ eV for $\bar N_d^+ = 5 \times 10^{18}$ cm$^{-3}$ under the self-averaged (nonlocal) impurity scattering. The figure on the right is a magnified plot of the region enclosed by the dashed line in the left figure. 
} 
  \label{fig:scmatrix_jelly}
 \end{center}
 \end{minipage}
 \end{figure}

Figure~\ref{fig:scmatrix_jelly} shows similar contour plots of the scattering-rate matrix under the {\em self-averaged impurity} of $\bar N_d^+ = 5 \times 10^{18}$ cm$^{-3}$. 
The self-averaged scattering-rate matrix is obtained from the quasi-1D version of Eq.~\eqref{eq:sv_sigma}.
It should be noted that the scattering-rate matrix is non-diagonal even under the self-averaged impurity because of the `center-of-mass' position dependence of the self-energy. Hereafter, such impurity scattering induced by non-diagonal scattering-rate matrices is referred to as `nonlocal' and the impurity scattering approximated by diagonal scattering-rate matrices, which is similar to the `jellium' impurity in the traditional approach, is referred to as `local.'

The scattering rate per unit length due to impurity scattering at $z$ ($=ma$) and $E$ is calculated by 
\begin{align}
\frac{1}{{\tau _{imp} \left( {ma;E} \right)}}   = \frac{1}{{a \hbar}} {\Gamma _{imp}}\left( {m,m;E} \right) ,
\label{eq:tau_{imp}}
\end{align}
and the transition rate per unit length due to the left (right) contact, $1/\tau _{L(R)}$, is given by the imaginary part of 
Eq.~\eqref{eq:1DSigmaLR^r} divided by $\hbar$. 
Since $1/\tau _{L(R)}$ is dependent on $a$, it is crucial to choose $a$ such that $1/\tau _{L(R)}$ is much larger than $1/\tau _{imp}$. Otherwise, the device region would be nearly closed at the contacts and no current would flow. 

\begin{figure}[tb]%
 \begin{minipage}{1\hsize}
 \begin{center}
  \includegraphics[width=8.7cm]{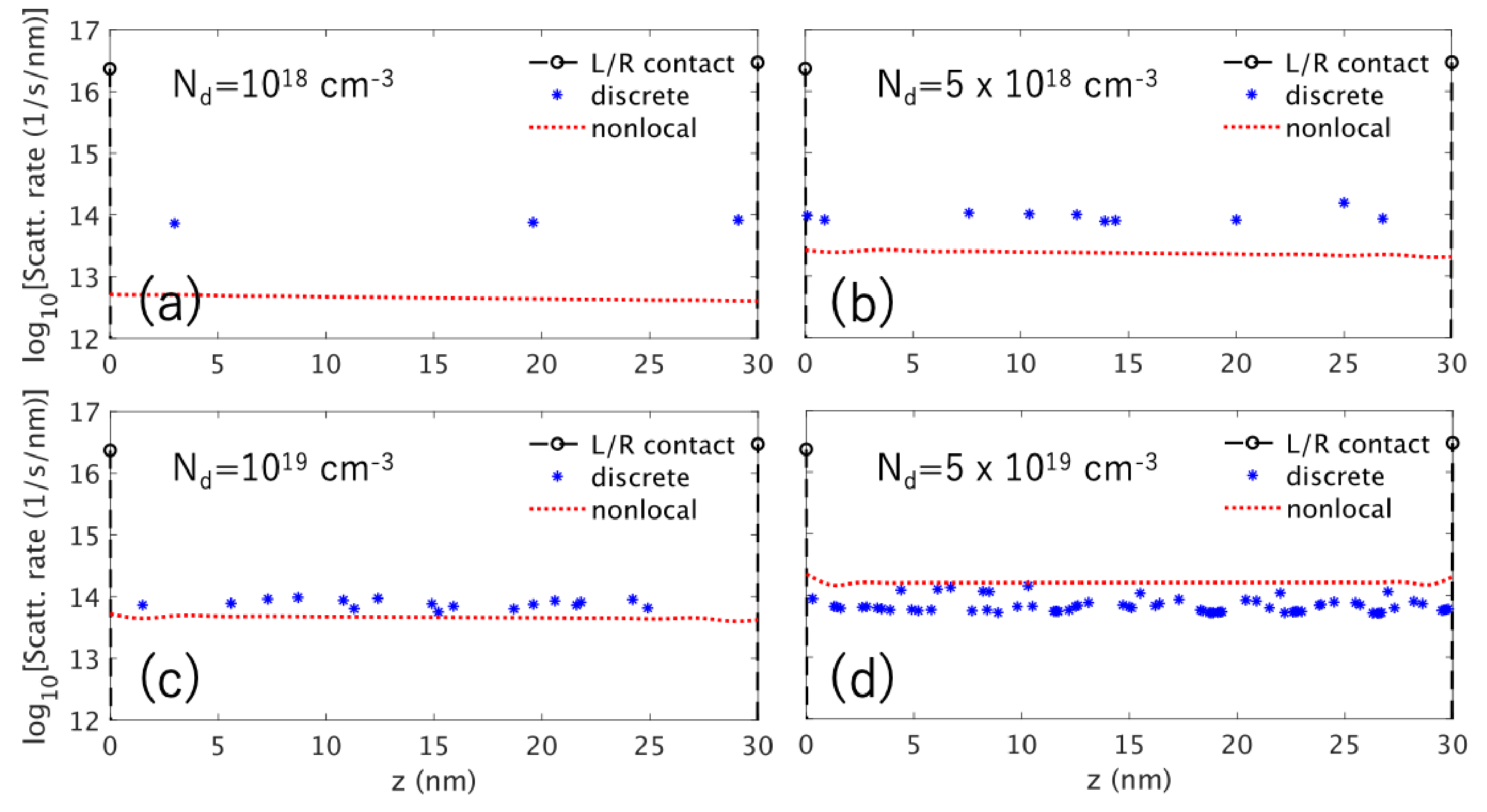}
  \caption{
  Scattering rates per unit length (blue symbols) at $E - {\varepsilon _{01}} = 0.05$ eV under (a) $\bar N_d^+ =10^{18}$, (b) $5 \times 10^{18}$, (c) $10^{19}$, and  (d) $5 \times 10^{19}$ cm$^{-3}$. The red dotted curves represent those under the self-averaged impurity (denoted as nonlocal) and the black dashed lines with open circles represent those at the left and right boundary (denoted as L/R contact).
} 
  \label{fig:sc_rate}
 \end{center}
 \end{minipage}
 \end{figure}

Figure~\ref{fig:sc_rate} shows the impurity scattering rates per unit length at $E - {\varepsilon _{01}} = 0.05$ eV with blue symbols under various impurity densities; $\bar N_d^+ =10^{18}$, $5 \times 10^{18}$, $10^{19}$, and $5 \times 10^{19}$ cm$^{-3}$. The scattering rates under the self-averaged impurity (denoted as nonlocal) are shown with red dotted curves and those at the left and right contacts (denoted as L/R contact) are shown with black dashed lines with open circles. 
In the present simulations, we have employed $a = 0.1$ nm, which is small enough to justify the present simulations; $1/\tau _{L(R)}$ is much larger than $1/\tau _{imp}$. 
We notice that in the regimes of ${\bar N}_{d}^{+} \le 10^{19}$ cm$^{-3}$, the scattering rates under the self-averaged impurity tend to be much weaker than those under the discrete impurities. This is quite reasonable because spatial locality of impurity scattering becomes more pronounced as ${\bar N}_{d}^{+}$ gets smaller. 
When $ {\bar N}_{d}^{+}$  is smaller than $10^{18}$ cm$^{-3}$, impurity scattering is usually less significant compared with phonon scattering~\cite{Jacoboni2010book,Fischetti2016book}. Therefore, the discrete nature of impurity scattering would be most significant when the impurity density is moderate ($10^{18} \le {\bar N}_{d}^{+} \le 10^{19}$ cm$^{-3}$). This is consistent with the previous finding from the Monte Carlo simulations~\cite{Sano2021}. On the other hand, as ${\bar N}_{d}^{+}$ is greater than $10^{19}$ cm$^{-3}$, the spatial distribution of impurities is averaged out and the scattering rates become close to those under the self-averaged impurity.

\begin{figure}[tb]%
 \begin{minipage}{1\hsize}
 \begin{center}
  \includegraphics[width=8.7cm]{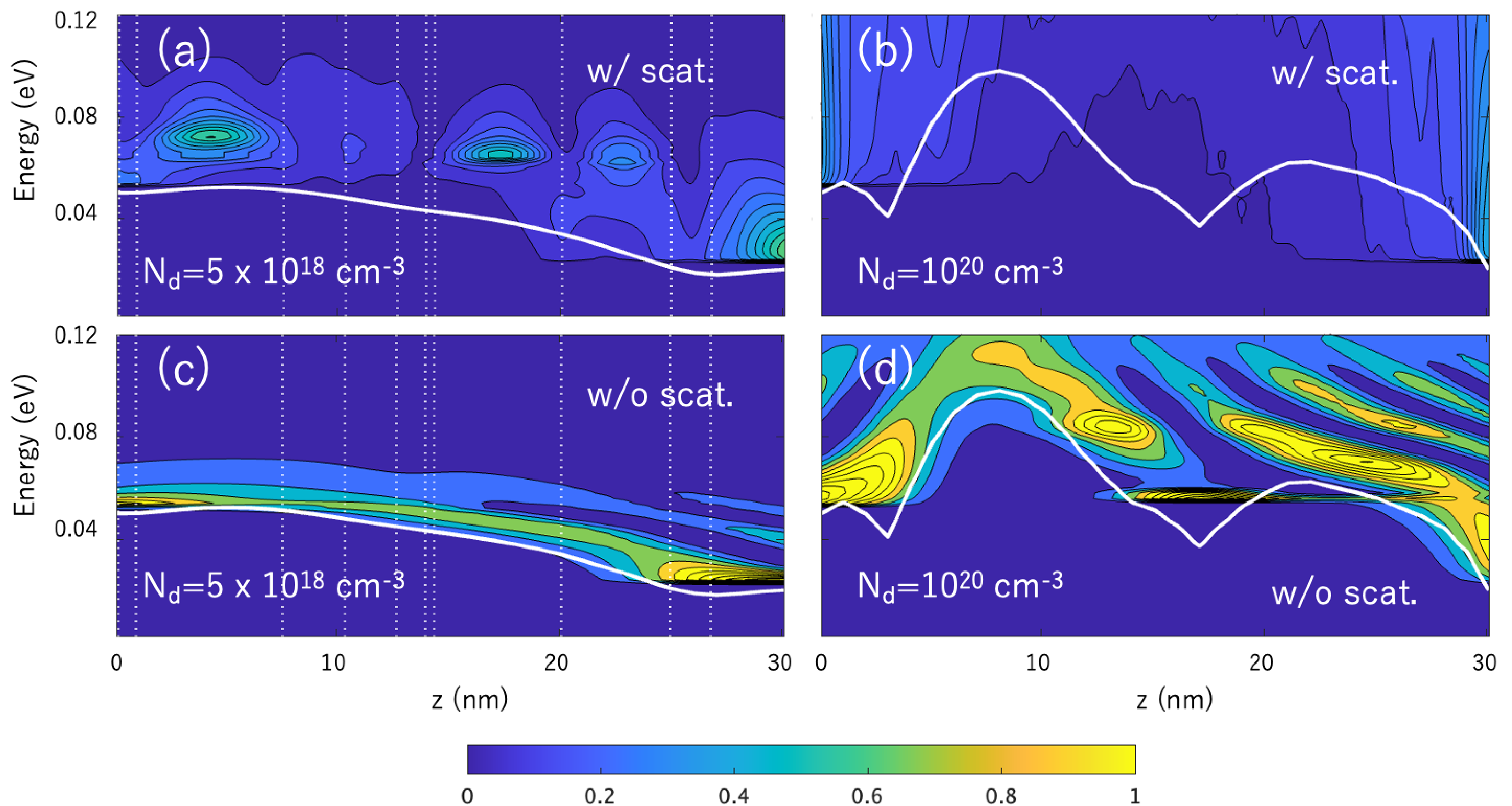}
  \caption{
  Energy spectra of the electron density under $\bar N_d^+ = 5 \times 10^{18}$  (left) and $10^{20}$ cm$^{-3}$ (right). The electron densities including the short-range impurity scattering are shown in (a) and (b), whereas those without impurity scattering are shown in (c) and (d). The white curves show the band-edge of the lowest subband and the vertical white dotted lines in (a) and (c) represent the positions of impurities.
} 
  \label{fig:ldos}
 \end{center}
 \end{minipage}
 \end{figure}

The energy spectra of the electron density under $\bar N_d^+ = 5 \times 10^{18}$ and $10^{20}$ cm$^{-3}$ with and without short-range impurity scattering are shown in Fig.~\ref{fig:ldos}. 
The electron density is calculated by Eq.~\eqref{eq:n_3d}. 
Notice that the long-range potential fluctuations induced by discrete impurities are represented by white solid curves. The vertical white dotted lines in (a) and (c) represent the positions of impurities.
When the short-range impurity scattering is turned off, the phase coherence is clearly observed under both (c) $\bar N_d^+ = 5 \times 10^{18}$ and (d) $10^{20}$ cm$^{-3}$. This reflects the fact that the electron transport is purely ballistic over the entire device region so that the phase interference is induced by  the `long-range' potential fluctuations. As a consequence, even a quasi-bound state in the potential dip at $z= 18$ nm in (d) is observed. 
As soon as impurity scattering is turned on, the electron density is smeared out and the phase coherence under high impurity density is completely diminished, as shown in (b). 
It is, however, quite interesting that as far as $\bar N_d^+$ is moderate, phase interference still remains in (a) {\em in the spaces between localized impurities}  even when the impurity scattering is active. This phenomenon may be observable at low temperatures in experiments in which phonon scattering is depressed.

\begin{figure}[tb]%
 \begin{minipage}{1\hsize}
 \begin{center}
  \includegraphics[width=8.7cm]{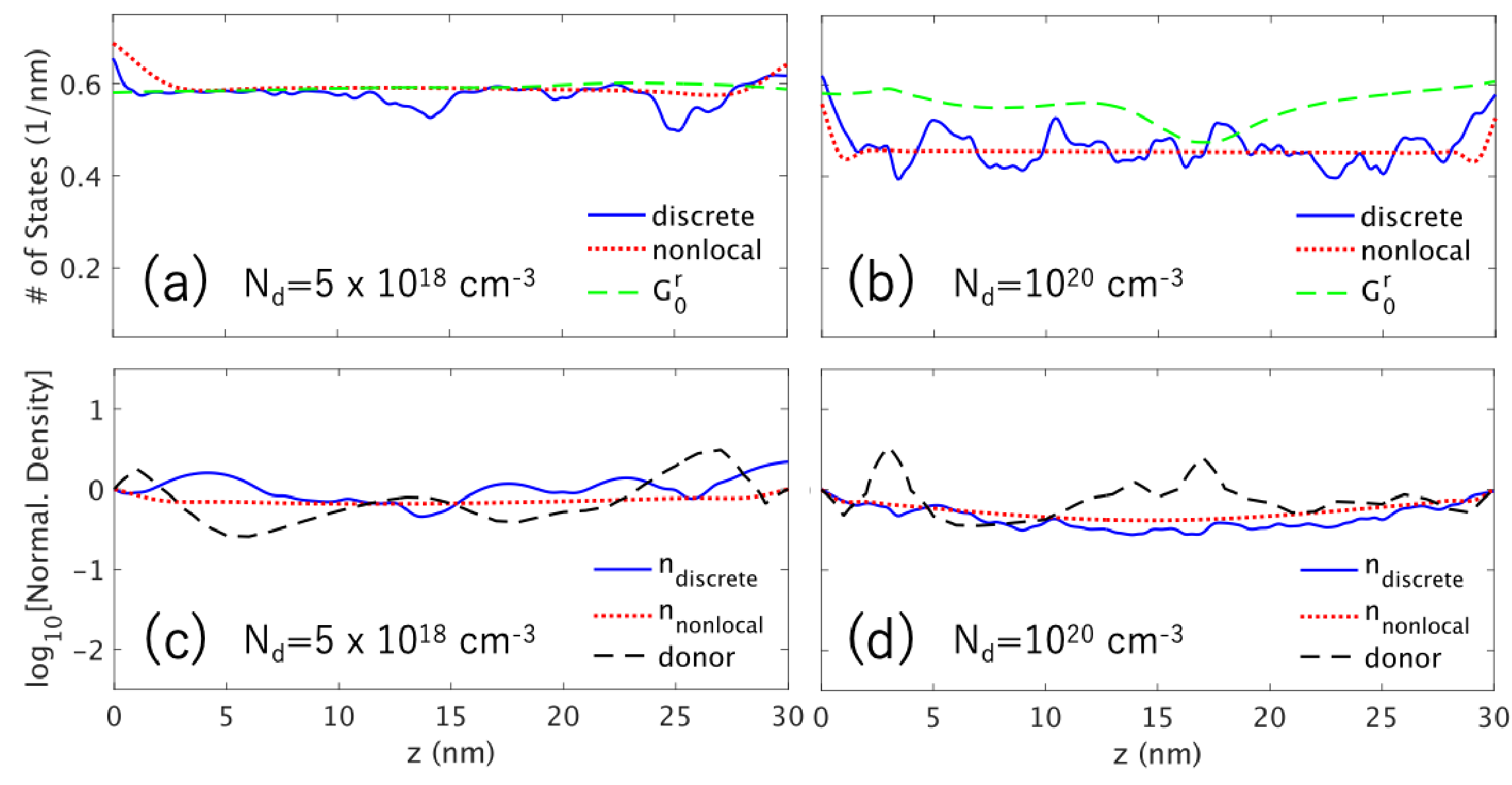}
  \caption{
  Number of states obtained from the LDOS by integrating over electron's energy from $E - {\varepsilon _{01}}=0$  to  
  0.4 eV for $\bar N_d^+ = 5 \times 10^{18}$ (left) and $10^{20}$ cm$^{-3}$ (right). The blue solid and red dotted curves show those under discrete impurities and the (nonlocal) self-averaged impurity, respectively.  
  The green dashed curves in (a) and (b) are those without short-range impurity scattering, and black dashed curves in (c) and (d) show the effective donor impurity densities.
 } 
  \label{fig:ldos_en}
 \end{center}
 \end{minipage}
 \end{figure}

Figure~\ref{fig:ldos_en} shows the number of states per unit length and the normalized electron densities as a function of position for $\bar N_d^+ = 5 \times 10^{18}$ and $10^{20}$ cm$^{-3}$. 
The number of states per unit length is calculated by integrating the LDOS from $E - {\varepsilon _{01}} =$ 0 to 0.4 eV. The electron density is normalized with the electron density at the left end ($z=0$ nm) so that it could be directly compared with the 3-D `long-range' donor density similarly normalized. 
The results under the nonlocal self-averaged impurities are also shown with red dotted curves. 
It is quite surprising that in spite of very different transport mechanisms, the results under the (nonlocal) self-averaged impurities well reproduce the averaged behavior of those under discrete impurities. Hence, the {\em ensemble-averaged} characteristics of electron transport in nanowires may be well approximated by the {\em nonlocal self-averaged} impurity scattering. 

\begin{figure}[tb]%
 \begin{minipage}{1\hsize}
 \begin{center}
  \includegraphics[width=6.5cm]{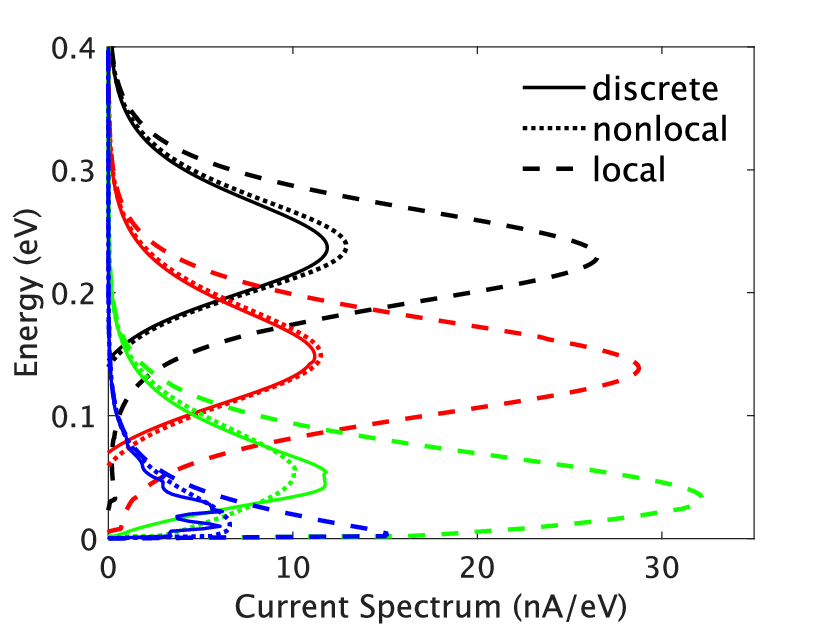}
  \caption{%
  Energy spectra of the current of the wires for 
{
$\bar N_d^+ = 10^{20}$ (black),~$5 \times 10^{19}$ (red),~ $10^{19}$ (green), and $10^{18}$ cm$^{-3}$ (blue).  The solid curves represent those under discrete impurities, whereas the dotted and dashed curves} are those under the nonlocal and local self-averaged impurity scattering, respectively.
} 
  \label{fig:Ispect}
 \end{center}
 \end{minipage}
 \end{figure}

This conjecture is supported by Fig.~\ref{fig:Ispect}, in which the energy spectra of the current under three different self-energies due to impurity scattering (discrete, nonlocal self-averaged, and local self-averaged) for $\bar N_d^+ = 10^{18},~10^{19},~ 5 \times 10^{19}$, and $10^{20}$ cm$^{-3}$ are shown. 
The shift of the current spectra toward the higher energy regions simply reflects the shift of $\mu _L$ in the left lead due to the increase of the electron density. 
Again, the nonlocal self-averaged impurity scattering reproduces well the current spectra under the discrete impurities except the regimes of low impurity density. In the low density regions, the current becomes strongly dependent of the spatial impurity distribution inside the wire and show large variations. Thus, the current under discrete impurities are not necessarily similar to the ensemble-averaged current. 
On the other hand, the local self-averaged impurity scattering tends to overestimate the current spectra. This is consistent with the fact that dephasing due to elastic scattering is weaker when the scattering-rate matrix is approximated by a diagonal matrix~\cite{Golizadeh-Mojarad2007}.  
Since the traditional treatment of impurity scattering is based on the local approximation, the usage of such impurity scattering should be with great caution especially when the impurity density is moderate ($10^{18} \le {\bar N}_{d}^{+} \le 10^{19}$ cm$^{-3}$) where the discrete nature of impurities is most significant. 

The total current and mobility from the present NEGF simulations by using three different self-energies due to impurity scattering are summarized in Table~\ref{tab:table1}. The current and mobility are evaluated from Eqs.~\eqref{eq:IcLR} and \eqref{eq:mob}, respectively. 
\begin{table}[tb]
\caption{%
Currents and mobilities of the wires of $\rho _{s} = 4$ nm with the discrete, nonlocal self-averaged, and local self-averaged impurity scattering under various donor densities.}
\label{tab:table1}
\begin{ruledtabular}
\begin{tabular}{lrrrr}
$\bar N_d^+$ (cm$^{-3}$) & 10$^{18}$ & 10$^{19}$ & 5 $\times$ 10$^{19}$ & 10$^{20}$\\
\hline
$I_{disc}$ ($\mu$A) & 0.129 & 0.300 & 0.319 & 0.311\\
$\mu_{disc}$ (cm$^2$/V/s) & 217 & 95 & 41 & 29\\
\\
$I_{nonloc}$ ($\mu$A) & 0.155 & 0.320 & 0.344 & 0.363\\
$\mu_{nonloc}$ (cm$^2$/V/s) & 229 & 100 & 51 & 41\\
\\
$I_{loc}$ ($\mu$A) & 0.264 & 1.18 & 1.29 & 1.18\\
$\mu_{loc}$ (cm$^2$/V/s) & 406 & 215 & 127 & 98\\
\end{tabular}
\end{ruledtabular}
\end{table}
In general, the values of discrete impurities in Table~\ref{tab:table1} greatly fluctuate even under the same $\bar N_d^+$ due to different spatial distributions of impurities in the substrate and, thus, they should not be regarded as typical values for the given impurity densities. More details of such statistical properties in electron transport by following the present approach have been reported in \cite{Sano2025}. 
Also, we observe that the local self-averaged scattering (the traditional approach) tends to overestimate the mobility compared with that in the nonlocal self-averaged scattering, as mentioned above. 

Finally, we would like to make a comment on the current obtained from the present NEGF simulations. The left and right current spectra, which are given by the integrand of $I_{L}$ and $I_{R}$ in Eq.~\eqref{eq:IcLR}, are supposed to be symmetric; their magnitudes should be equal under steady state. This is indeed true in the present scheme when impurity density is moderate, say, $\bar N_{d}^{+} \le 10^{19}$ cm$^{-3}$. However, they become asymmetric as $\bar N_{d}^{+}$ increases further. This may be due to the lack of inelastic phonon scattering in the present NEGF simulations and the lack of the self-consistent loop between Poisson's equation and the NEGF calculations. Fully self-consistent 3-D simulations between the NEGF including inelastic phonon scattering and Poisson's equation are certainly desired for more quantitative analyses and left for future study.

\section{\label{sec:concl}Conclusions}
We have proposed a new theoretical framework of the NEGF scheme to account for the discrete nature of impurities in electron transport  under semiconductor nanostructures. The electrostatic potential of each impurity is separated into the long- and short-range parts, and implemented into both the Poisson equation and the NEGF in a self-consistent manner. 
The self-energy treats the short-range part of the impurity potential due to impurity scattering, whereas the long-range part of the potential is treated as the self-consistent Hartree potential. We have succeeded in deriving the (global) {\em position-dependent} impurity scattering rate systematically {\em under inhomogeneous impurity profiles} and in clarifying its physical meaning. Although the formula is nearly identical to the one found in the literature from Fermi's golden rule, the position dependence of the scattering rate is represented by the Wigner coordinates, rather than the ordinary real-space coordinates. As a result, the impurity scattering becomes intrinsically nonlocal in space. Furthermore, by imposing the average over impurity distributions (`self-averaging'), the present scheme properly reproduces the traditional Green's function approach with the self-energy due to `jellium' impurity.

The present theoretical scheme has been applied to cylindrical wires of $\rho _{s} =$ 4 nm and $L_{c}=30$ nm under the quasi-1D approximation. It has been shown how the discrete nature of impurities  is represented in the scattering-rate matrices, electron density, and transport characteristics such as current and mobility. Especially, the scattering-rate matrix is intrinsically non-diagonal, and the dephasing effects due to impurity scattering are stronger than those in the traditional approaches where the scattering-rate matrix is strictly diagonal. As a result, the current and mobility would be overestimated by the traditional schemes. 
In addition, we have found that the {\em ensemble-averaged} transport properties such as current and mobility may be well approximated by the {\em nonlocal self-averaged} impurity scattering. 

{
We would also like to point out that the present theoretical framework of the NEGF can be easily applied to other scattering mechanisms and, thus, it is quite interesting to investigate how the nonlocality shows up in those scattering processes. 
Since the nonlocality is associated with the Wigner coordinates, we expect that it would be significant, especially in the polar electron-phonon scattering and roughness scattering.}

\begin{acknowledgments}
The author wishes to acknowledge the support by JSPS KAKENHI under Grant JP23K22809.
\end{acknowledgments}



%

\end{document}